\documentclass[reprint,superscriptaddress]{revtex4-2}

\usepackage{graphicx}
\usepackage{amsmath,amsfonts,amssymb,amsthm,color}
\usepackage[normalem]{ulem}  
\usepackage{multirow}
\usepackage{booktabs} 
\allowdisplaybreaks

\usepackage{hyperref}
\hypersetup{
    colorlinks = true,
    urlcolor   = blue,
    citecolor  = black,
}

\newcommand{\RomanNumeralCaps}[1]

\usepackage[linesnumbered,ruled,vlined]{algorithm2e}

\begin{document}

\title{Detecting hidden structures from a static loading experiment: \\ topology optimization meets physics-informed neural networks}

\author{Saviz Mowlavi}
 \email{mowlavi@merl.com}
\affiliation{Mitsubishi Electric Research Laboratories, Cambridge, MA 02139, USA}
\affiliation{Department of Mechanical Engineering, MIT, Cambridge, MA 02139, USA}
\author{Ken Kamrin}%
 \email{kkamrin@mit.edu}
\affiliation{Department of Mechanical Engineering, MIT, Cambridge, MA 02139, USA}
\affiliation{Department of Mechanical Engineering, UC Berkeley, Berkeley, CA 94720, USA}

\begin{abstract}
Most noninvasive imaging techniques utilize electromagnetic or acoustic waves originating from multiple locations and directions to identify hidden geometrical structures. 
Surprisingly, it is also possible to image hidden voids and inclusions buried within an object using a single static thermal or mechanical loading experiment by observing the response of the exposed surface of the body, but this problem is challenging to invert. 
Although physics-informed neural networks (PINNs) have shown promise as a simple-yet-powerful tool for problem inversion, they have not yet been applied to \textcolor{black}{imaging} problems with a priori unknown topology. 
Here, we introduce a topology optimization framework based on PINNs that identifies concealed geometries using exposed surface data from a single loading experiment, without prior knowledge of the number or types of shapes. 
We allow for arbitrary solution topology by representing the geometry using a material density field combined with a novel eikonal regularization technique. 
We validate our framework by detecting the number, locations, and shapes of hidden voids and inclusions in many example cases, in both 2D and 3D, and we demonstrate the method's robustness to noise and sparsity in the data. 
Our methodology opens a pathway for PINNs to solve \textcolor{black}{geometry optimization problems in engineering}.
\end{abstract}

\maketitle

Noninvasive detection of hidden geometries is desirable in countless applications including medical imaging and diagnosis \citep{suetens2017}, nondestructive evaluation of materials \citep{hellier2013}, or mine detection \cite{robledo2009}. Common noninvasive imaging methods, such as computed tomography \citep{withers2021}, ultrasonic testing \citep{krautkramer2013}, and magnetic resonance imaging \citep{mcrobbie2017}, rely upon electromagnetic or acoustic waves sent from multiple directions and positions to detect the locations and shapes of structures hidden inside a given body \citep{kasban2015}.
These noninvasive imaging techniques often require costly specialized equipment and long data acquisition procedures, limiting their applicability in cost- or time-sensitive applications \citep{de2014}.

It is known that \textcolor{black}{for} a two-dimensional \textcolor{black}{(2D)} elastic body containing one void, a \textit{single} mechanical experiment provides sufficient data to uniquely identify the shape and location of the void, by utilizing the physics governing the problem \citep{ang1999}. In practice, this involves subjecting the body to a boundary traction, measuring the resulting surface displacement over a finite portion of the boundary, and solving a physics-constrained inverse problem for the geometry of the void. We show here that this uniqueness result can, remarkably, be generalized to any number of voids or solid inclusions in three-dimensional \textcolor{black}{(3D)} bodies, and any type of experiment where the physics is governed by linear or nonlinear elliptic partial differential equations (PDEs), such as heat conduction or large-deformation elasticity. This opens the door to a new noninvasive imaging paradigm where buried voids or inclusions are detected from surface measurements acquired in a single loading experiment, with the potential to accelerate the data acquisition process in time-critical applications. However, the inverse problem is extremely challenging to solve due to the unknown topology, the limited amount of measurements, and the complexity of the underlying physical laws \citep{bonnet2005,colton2019}.

In the present paper, we introduce a novel and simple topology optimization (TO) method that leverages physics-informed neural networks (PINNs) for accurately reconstructing the shapes and locations of concealed objects using surface data obtained from a single experiment along with knowledge of the physics. Previous attempts to solve related inverse problems rely on more complicated methods such as adjoint-based optimization
and do not always yield satisfactory results, especially in the presence of sparse measurements acquired by only one or a few sets of experiments \citep{lee2000,ameur2004,mei2016,mei2021}. 

PINNs have emerged in recent years as a robust tool for problem inversion across disciplines and over a range of model complexity \citep{dissanayake1994,lagaris1998,raissi2019,sun2020}. Their ability to seamlessly blend measurement data or design objectives with governing PDEs in nontrivial geometries has enabled practitioners to solve easily a range of inverse problems involving identification or design of unknown properties in fields ranging from mechanics to optics and medicine \citep{raissi2020,sahli2020,lu2021,haghighat2021,chen2022,mowlavi2023,joglekar2023dmf,jeong2023complete,qiu2023}. Encouraged by these early successes, we adopt PINNs as the foundation of our TO framework for geometry detection. Building on the strength of PINNs, our approach leverages both the measurements and the governing PDEs, is straightforward to implement regardless of the complexity of the physical model, and produces accurate results without requiring a training dataset, unlike purely data-based machine learning approaches to solve related inverse problems \citep{ni2021,yang2023,crocker2023}. To the best of our knowledge, the present work marks the first time that PINNs have been applied to \textcolor{black}{imaging} problems involving \textit{a priori} unknown topology and geometry.

Using data from a single loading experiment, our PINN-based TO framework identifies complex concealed shapes without any prior knowledge on the number and types of shapes. We allow for arbitrary solution topology by representing the geometry using a material density field equal to 0 in one phase and 1 in the other. The material density is parameterized through a neural network, which needs to be regularized in order to push the material density towards 0 or 1 values. Thus, one key ingredient in our framework is a novel eikonal regularization, inspired from fast-marching level-set methods \citep{adalsteinsson1995,osher2004} and neural signed distance functions \citep{gropp2020}, that promotes a constant thickness of the interface region where the material density transitions between 0 and 1, leading to well-defined boundaries throughout the domain. 
This eikonal regularization enters as an additional term in the standard PINN loss, which is then used to train the neural networks underlying the material density and physical quantities to yield a solution to the geometry detection problem. 

As an illustration, we apply our framework to \textcolor{black}{2D and 3D} cases involving elastic bodies under mechanical \textcolor{black}{or} thermal loading, and discover the topology, locations, and shapes of hidden structures for a variety of geometries and materials. Finally, comparisons with other regularization methods commonly employed in TO show that the eikonal regularization consistently yields the most accurate results in these applications.

\section*{Results} 

\subsection*{Problem formulation}

We consider noninvasive geometry detection problems of the following \textcolor{black}{general} form. Suppose we have a continuous body $\mathcal{B}$ \textcolor{black}{in 2D or 3D} containing an unknown number of hidden voids or inclusions, with unknown shapes and at unknown locations within the body. The material properties are assumed to be known and homogeneous within the body and the inclusions. We then apply a certain type of loading (e.g.~mechanical or thermal) on the body's external boundary $\partial \mathcal{B}^\mathrm{ext}$, which produces a response within the body that can be described by a set of $n$ physical quantities (e.g.~displacements, stresses, temperature, etc). These physical quantities can be lumped into a vector field $\boldsymbol{\psi} : \mathcal{B} \rightarrow \mathbb{R}^n$ and satisfy a known set of governing PDEs \textcolor{black}{and applied boundary conditions (BCs)}. The goal of the inverse problem is to identify the number, locations, and shapes of the voids or inclusions based on measurements \textcolor{black}{of the boundary data that arises along a finite connected segment} of $\partial \mathcal{B}^\mathrm{ext}$. The $\boldsymbol{\psi}$ output one measures on the boundary is the data ``conjugate'' to the prescribed BCs; e.g.~if the problem were elastic and surface stresses were applied one would measure the surface displacements that emerge, or if the problem were thermal and surface temperatures were prescribed one would measure the resulting normal heat flux.

\textcolor{black}{\textbf{Theorem.} As long as the governing PDEs are elliptic, this inverse problem has a unique solution. The proof is provided in the Supplementary Materials. This uniqueness result provides an essential motivation for the PINN-based TO framework that we introduce in the next section.}

As a concrete example, consider two \textcolor{black}{2D prototypical} plane-strain elasticity inverse problems.
\begin{figure}[tb]
\centering
\includegraphics[width=0.5\textwidth]{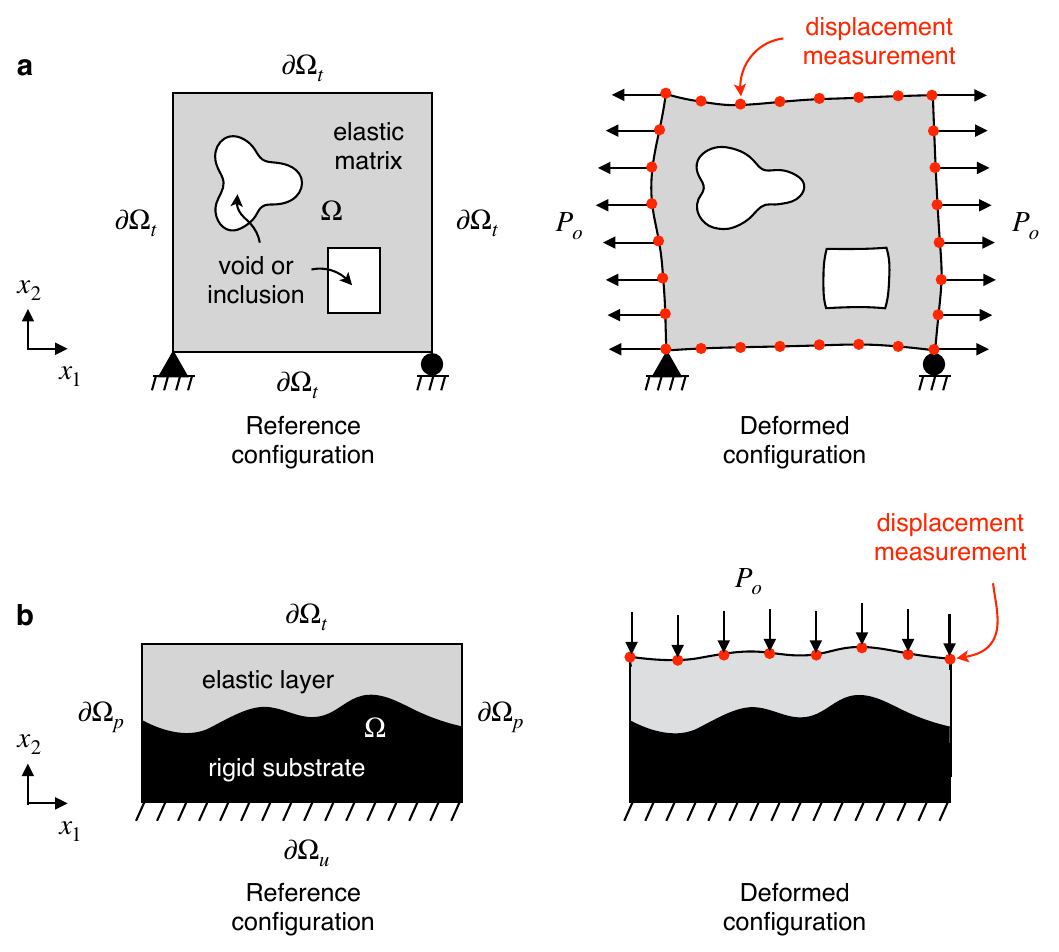}
\caption{\textbf{Setup of two \textcolor{black}{prototypical} geometry identification problems in elastic bodies under mechanical loading.} \textbf{a}, A square elastic matrix with hidden voids or inclusions is pulled by a known uniform traction on two opposite sides. The goal is to identify the number, locations, and shapes of the voids or inclusions within using measurements of the displacement occurring along the outer boundary of the matrix. \textbf{b}, An elastic layer on top of a hidden rigid substrate is compressed from the top by a uniform pressure. The goal is to identify the shape of the substrate using measurements of the displacement of the top surface.}
\label{fig:Geometry}
\end{figure}
In the first case, a square elastic matrix with hidden voids or inclusions is pulled by a uniform traction $P_o$ on two sides (Fig.~\ref{fig:Geometry}a). The goal of the inverse problem is to identify the number, locations, and shapes of the voids or inclusions using discrete measurements of the displacement of the outer boundary of the matrix. In the second case, an elastic layer on top of a hidden rigid substrate is compressed from the top by a uniform pressure $P_o$, with periodic lateral BCs (Fig.~\ref{fig:Geometry}b). The goal is to identify the shape of the substrate using discrete measurements of the displacement of the top surface. For both cases, \textcolor{black}{the governing PDEs, applied BCs, and} the constitutive properties of all materials are assumed to be known. We consider two different types of constitutive laws: compressible linear elasticity, which characterizes the small deformation of any compressible elastic material, and incompressible nonlinear hyperelasticity, which models the large deformation of rubber-like materials. \textcolor{black}{We also presume that the discrete measurements are closely distributed in order to capture the continuous displacement field along the corresponding surface.}

\textcolor{black}{As in} density-based TO methods \citep{sigmund2013}, we avoid any restriction on the number and shapes of hidden structures by parameterizing the geometry of the elastic body $\mathcal{B}$ through a discrete-valued material density function $\rho : \Omega \rightarrow \{0,1\}$, where $\Omega$ is a global domain comprising both $\mathcal{B}$ and the hidden voids or inclusions. The material density is defined to be equal to 1 in the elastic body $\mathcal{B}$ and 0 in the voids or inclusions. 
The physical quantities $\boldsymbol{\psi}$ can then be extended to the global domain $\Omega$ by introducing an explicit $\rho$-dependence in their governing PDEs, leading to equations of the form 
\begin{subequations}
\begin{align}
\mathbf{r}(\boldsymbol{\psi}(\mathbf{x}),\rho(\mathbf{x})) &= 0, \quad \mathbf{x} \in \Omega, \label{eq:GoverningPDEs} \\
\mathbf{b}(\boldsymbol{\psi}(\mathbf{x})) &= 0, \quad \mathbf{x} \in \partial \Omega, \label{eq:AppliedBCs}
\end{align}
\end{subequations}
\textcolor{black}{where $\mathbf{r}$ contains the PDE residuals and $\mathbf{b}$ encodes the applied} BCs defined on the external boundary $\partial \Omega = \partial \mathcal{B}^\mathrm{ext}$. \textcolor{black}{Both} $\mathbf{r}$ and $\mathbf{b}$ may contain partial derivatives of $\boldsymbol{\psi}$ and $\rho$. The inverse problem is now to find the distribution of material density $\rho$ in $\Omega$ so that the corresponding solution for $\boldsymbol{\psi}$ matches surface measurements $\boldsymbol{\psi}_i^m$ at discrete locations $\mathbf{x}_i \in \partial \Omega^m \subset \partial \Omega$, that is,
\begin{equation}
\boldsymbol{\psi}(\mathbf{x}_i) = \boldsymbol{\psi}_i^m, \quad \mathbf{x}_i \in \partial \Omega^m.
\label{eq:Measurements}
\end{equation}
In practice, we might only measure select quantities in $\boldsymbol{\psi}$ at some of the locations, but we do not write so explicitly to avoid overloading the notation.

For the \textcolor{black}{prototypical} elasticity problems considered as an example \textcolor{black}{in Fig.~\ref{fig:Geometry}}, $\boldsymbol{\psi} = (\mathbf{u},\boldsymbol{\sigma})$ where $\mathbf{u}(\mathbf{x})$ and $\boldsymbol{\sigma}(\mathbf{x})$ are displacement and stress fields, respectively, and the governing PDEs \textcolor{black}{\eqref{eq:GoverningPDEs}} comprise equilibrium relations $\sum_j\partial \sigma_{ij}/\partial x_j = 0$ and a constitutive law $F(\boldsymbol{\sigma}, \nabla \mathbf{u}, \rho) = 0$, both defined over $\Omega$. The presence of $\rho$ in the constitutive law specifies different material behaviors \textcolor{black}{between} the elastic solid phase and the void or rigid inclusion phase. 
The applied BCs \textcolor{black}{\eqref{eq:AppliedBCs}} take the form $\boldsymbol{\sigma} \mathbf{n} = \bar{\mathbf{t}}$ on boundary segments $\partial \Omega_t$ subject to applied tractions ($\mathbf{n}$ is the outward unit normal), $\mathbf{u} = \bar{\mathbf{u}}$ on boundary segments $\partial \Omega_u$ subject to applied displacements, and periodic conditions for $\mathbf{u}$ and $\boldsymbol{\sigma}$ on boundary segments $\partial \Omega_p$. \textcolor{black}{Last, since surface displacements are measured, the constraint \eqref{eq:Measurements}} is $\mathbf{u}(\mathbf{x}_i) = \mathbf{u}_i^m$ for $\mathbf{x}_i \in \partial \Omega^m$. 

\textcolor{black}{Finally, we relax the binary constraint on} the material density \textcolor{black}{$\rho$ by allowing intermediate values} between 0 and 1, \textcolor{black}{which} renders the problem amenable to gradient-based optimization. \textcolor{black}{However, this requires adding an appropriate} regularization mechanism \textcolor{black}{to promote convergence of $\rho$} towards 0 \textcolor{black}{or} 1, \textcolor{black}{a key yet challenging task \citep{benning2018, bertero2021}.} As we will show in the discussion, common strategies employed in \textcolor{black}{computational imaging and} TO \citep{dorn2006,sigmund2013} do not yield satisfactory results in our PINN-based framework. Thus, we have developed a novel eikonal regularization scheme that we will describe after the general framework.

\subsection*{General \textcolor{black}{PINN-based TO} framework}
\label{sec:GeneralFramework}

\begin{figure*}
\centering
\includegraphics[width=\textwidth]{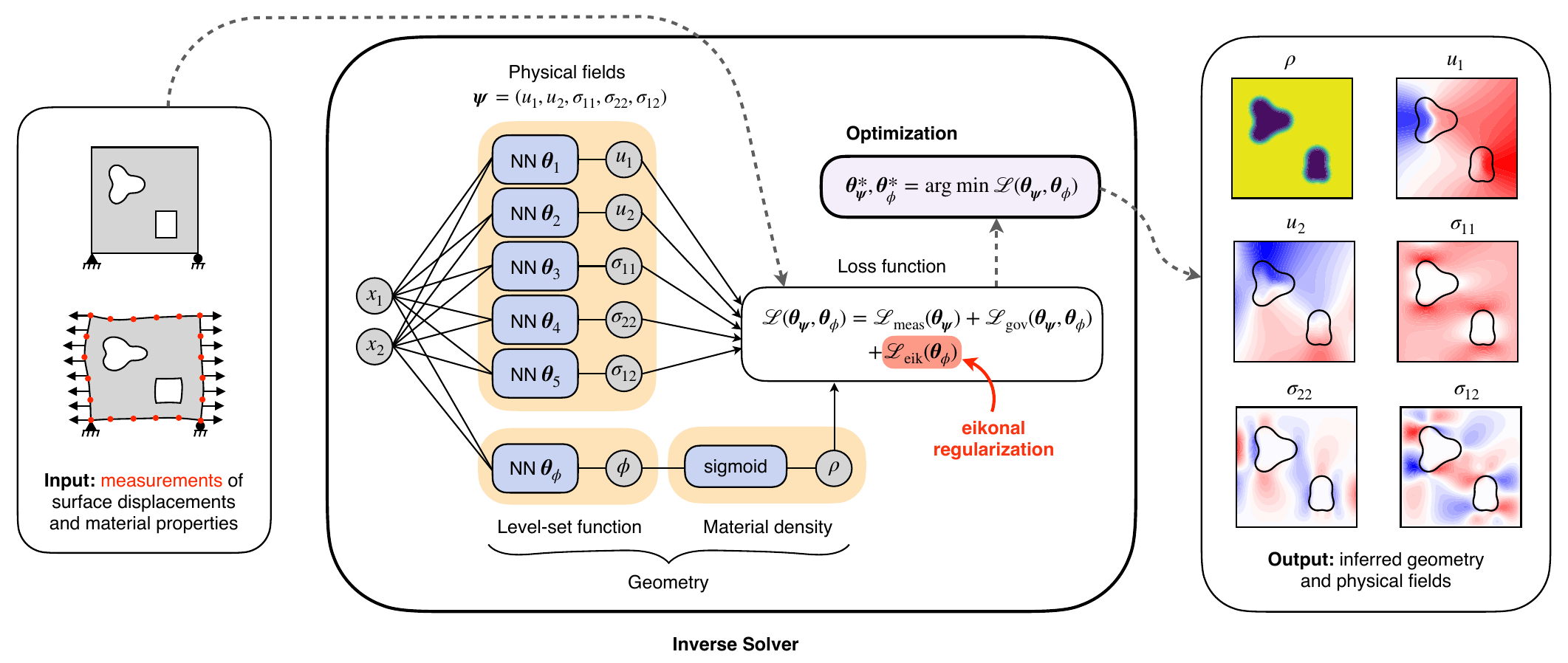}
\caption{\textbf{TO framework for noninvasive detection of hidden geometries.} The geometry of the system, which is initially unknown, is parameterized by a material density field given through a level-set function and equal to 1 in the elastic body and 0 in the voids or inclusions. The level-set function and the physical quantities describing the problem are approximated with deep neural networks designed to inherently satisfy the applied BCs. These neural networks are then trained to minimize a loss function that drives the material density and physical quantities towards satisfying the governing equations of the problem while matching discrete surface measurements. A crucial eikonal regularization term in the loss function ensures that the material density transitions between 0 and 1 over a prescribed length scale and avoids settling on intermediate values. By the end of the optimization, the converged material distribution reveals the location and shapes of the hidden structures.}
\label{fig:Framework}
\end{figure*}

We propose a TO framework based on PINNs for solving noninvasive geometry detection problems (Fig.~\ref{fig:Framework}). At the core of the framework are several deep neural networks that approximate the physical quantities $\boldsymbol{\psi}(\mathbf{x})$ describing the problem and the material density $\rho(\mathbf{x})$. For the physical quantities, each neural network maps the spatial location $\mathbf{x} = (x_1,x_2)$ to one of the variables in $\boldsymbol{\psi} = (\psi_1, \cdots, \psi_n)$; this can be expressed as $\psi_i = \bar{\psi}_i(\mathbf{x}; \boldsymbol{\theta}_i)$ where $\bar{\psi}_i$ is the map defined by the $i$th neural network and its trainable parameters $\boldsymbol{\theta}_i$ (see \textcolor{black}{SI}). For the material distribution, we first define a neural network with trainable parameters $\boldsymbol{\theta}_\phi$ that maps $\mathbf{x}$ to a scalar variable $\phi = \bar{\phi}(\mathbf{x}; \boldsymbol{\theta}_\phi)$. A sigmoid function is then applied to $\phi$ to yield $\rho = \mathrm{sigmoid}(\phi/\delta) = \mathrm{sigmoid}(\bar{\phi}(\mathbf{x}; \boldsymbol{\theta}_\phi)/\delta)$, which we simply write as $\rho = \bar{\rho}(\mathbf{x},\boldsymbol{\theta}_\phi)$. This construction ensures that the material density $\rho$ remains between 0 and 1, and $\delta$ is a transition length scale that we will comment on later. We define the phase transition to occur at $\rho = 0.5$ so that the zero level-set of $\phi$ delineates the boundary between the two material phases, hence $\phi$ is hereafter referred to as a level-set function \citep{osher1988,osher2004}.

We now seek the parameters $\boldsymbol{\theta}_{\boldsymbol{\psi}} = \{\boldsymbol{\theta}_1, \dots, \boldsymbol{\theta}_n\}$ and $\boldsymbol{\theta}_\phi$ so that the neural network approximations for $\boldsymbol{\psi}(\mathbf{x})$ and $\rho(\mathbf{x})$ satisfy the governing PDEs \eqref{eq:GoverningPDEs} and applied BCs \eqref{eq:AppliedBCs} while matching the surface measurements \eqref{eq:Measurements}. This is achieved by constructing a loss function of the form
\begin{align}
\mathcal{L}(\boldsymbol{\theta}_{\boldsymbol{\psi}}, \boldsymbol{\theta}_\phi) &= \lambda_\mathrm{meas} \mathcal{L}_\mathrm{meas}(\boldsymbol{\theta}_{\boldsymbol{\psi}}) + \lambda_\mathrm{gov} \mathcal{L}_\mathrm{gov}(\boldsymbol{\theta}_{\boldsymbol{\psi}}, \boldsymbol{\theta}_\phi) \nonumber \\ 
&\quad + \lambda_\mathrm{reg} \mathcal{L}_\mathrm{eik}(\boldsymbol{\theta}_\phi), \label{eq:TotalLoss}
\end{align}
where $\mathcal{L}_\mathrm{meas}$ and $\mathcal{L}_\mathrm{gov}$ measure the degree to which the neural network approximations do not satisfy the measurements and governing PDEs, respectively, $\mathcal{L}_\mathrm{eik}$ is a crucial regularization term that drives $\rho$ towards 0 or 1 values and that we will explain below, and the $\lambda$'s are scalar weights. The measurement loss takes the form
\begin{equation}
\mathcal{L}_\mathrm{meas}(\boldsymbol{\theta}_{\boldsymbol{\psi}}) = \frac{1}{|\partial \Omega^m|} \sum_{\mathbf{x}_i \in \partial \Omega^m} |\bar{\boldsymbol{\psi}}(\mathbf{x}_i;\boldsymbol{\theta}_{\boldsymbol{\psi}})-\boldsymbol{\psi}_i^m|^2,
\end{equation}
where $|\partial \Omega^m|$ denotes the size of the set $\partial \Omega^m$. A trivial modification of this expression is necessary in the case where only select quantities in $\boldsymbol{\psi}$ are measured. The PDE loss takes the form
\begin{equation}
\mathcal{L}_\mathrm{gov}(\boldsymbol{\theta}_{\boldsymbol{\psi}}, \boldsymbol{\theta}_\phi) = \frac{1}{|\Omega^d|} \sum_{\mathbf{x}_i \in \Omega^d} |\mathbf{r}(\bar{\boldsymbol{\psi}}(\mathbf{x}_i;\boldsymbol{\theta}_{\boldsymbol{\psi}}),\bar{\rho}(\mathbf{x}_i;\boldsymbol{\theta}_\phi))|^2,
\label{eq:LossGov}
\end{equation}
where $\Omega^d$ is a set of collocation points in $\Omega$, and we use automatic differentiation to calculate in a mesh-free fashion the spatial derivatives contained in $\mathbf{r}$. We design the architecture of our neural networks in such a way that they inherently satisfy the \textcolor{black}{applied} BCs (see \textcolor{black}{Supplementary Materials}).

Finally, the optimal parameters $\boldsymbol{\theta}_{\boldsymbol{\psi}}^*$ and $\boldsymbol{\theta}_\phi^*$ that solve the problem can be obtained by training the neural networks to minimize the loss \eqref{eq:TotalLoss} using stochastic gradient descent-based optimization. The corresponding physical quantities $\bar{\boldsymbol{\psi}}(\mathbf{x};\boldsymbol{\theta}_{\boldsymbol{\psi}}^*)$ will match the discrete surface measurements while satisfying the governing equations of the problem, while the corresponding material density $\bar{\rho}(\mathbf{x};\boldsymbol{\theta}_\phi^*)$ will reveal the number, locations, and shapes of the hidden voids or inclusions.

\subsection*{\textcolor{black}{Eikonal} regularization}
\label{sec:MaterialDensityRegularization}

We now describe the key ingredient that ensures the success of our framework. As mentioned above, the main challenge is to promote the material density $\rho(\mathbf{x})$ to converge towards 0 or 1 away from the material phase boundaries defined by the zero level-set $\phi = 0$. Moreover, we desire the thickness of the transition region along these boundaries, where $\rho$ goes from 0 to 1, to be uniform everywhere in order to ensure consistency of physical laws across the interface (e.g.~stress jumps). 

\begin{figure}
\centering
\includegraphics[width=0.5\textwidth]{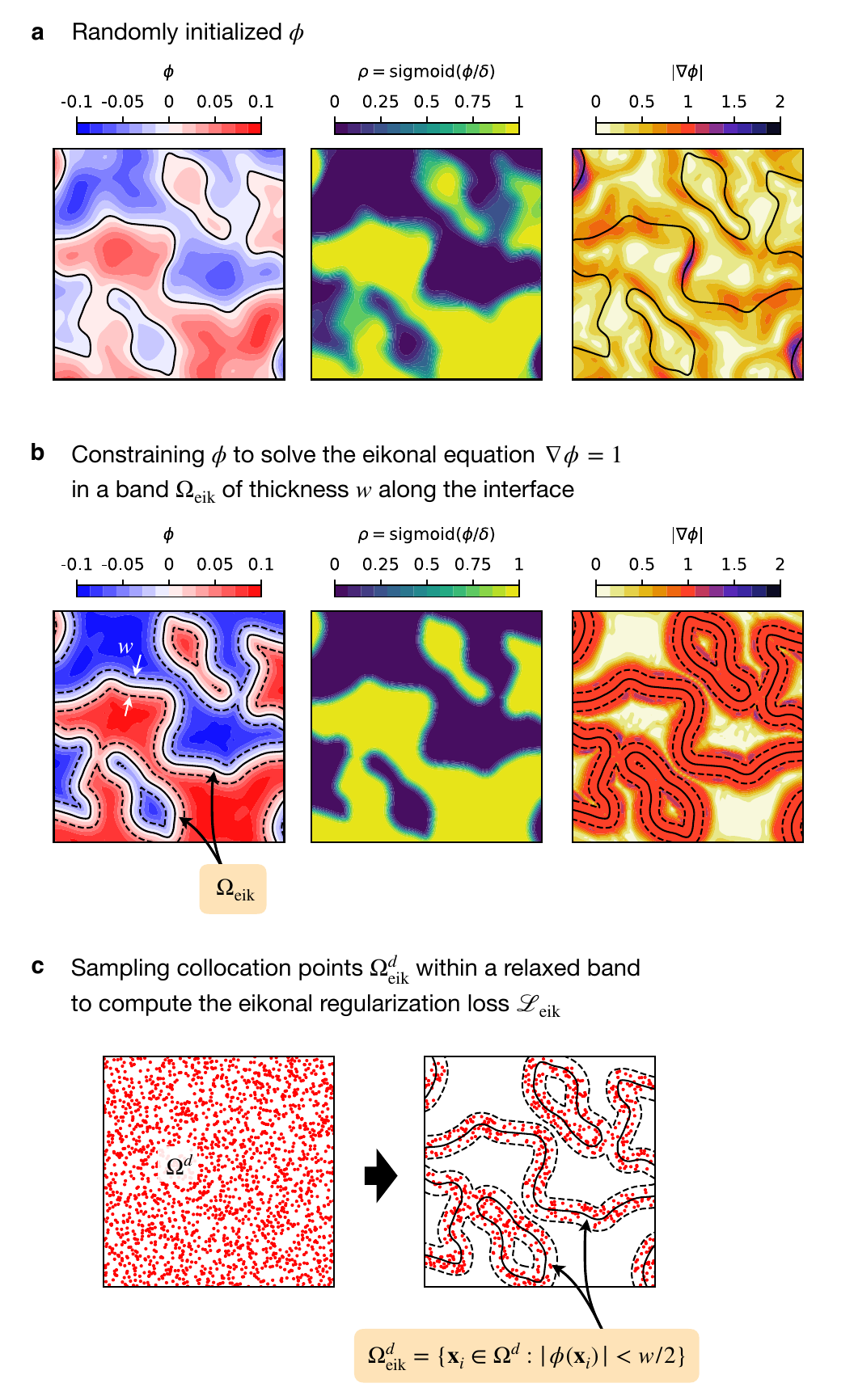}
\caption{\textbf{Eikonal regularization of the material density.} $\mathbf{a}$, A random level-set function $\phi$ yields a material density $\rho = \mathrm{sigmoid}(\phi/\delta)$ with large regions of values between 0 and 1, due to the nonuniformity of the gradient $|\nabla \phi|$ along the material boundaries defined by the zero level-set of $\phi$ (black lines). $\mathbf{b}$, Constraining $\phi$ to solve the eikonal equation $\nabla \phi = 1$ in a narrow band $\Omega_\mathrm{eik}$ of thickness $w$ (delineated by the dashed lines) along the material boundaries results in a uniform transition thickness of $\rho$ from 0 to 1, without large regions of intermediate density values. $\mathbf{c}$, The loss $\mathcal{L}_\mathrm{eik}$ implements the eikonal regularization in the PINN-based TO framework by penalizing violations of the constraint $|\nabla \phi| = 1$ in a subset of collocation points $\Omega_\mathrm{eik}^d \subset \Omega_d$ that approximates the true narrow band $\Omega_\mathrm{eik}$.}
\label{fig:LevelSet}
\end{figure}

To visualize what happens in the absence of regularization, consider a random instance of the neural network $\phi = \bar{\phi}(\mathbf{x},\boldsymbol{\theta}_\phi)$ (Fig.~\ref{fig:LevelSet}a, left) and the corresponding material distribution $\rho = \mathrm{sigmoid}(\phi/\delta)$ with $\delta = 0.01$ (Fig.~\ref{fig:LevelSet}a, center). The sigmoid transformation ensures that $\rho$ never drops below 0 or exceeds 1, leading to large regions corresponding to one phase or the other. However, the thickness of the transition region where $\rho$ goes from 0 to 1 is not everywhere uniform, resulting in large zones where $\rho$ assumes nonphysical values between 0 and 1 (Fig.~\ref{fig:LevelSet}a, center). This behavior stems from the non-uniformity of the gradient norm $|\nabla \phi|$ along the material boundaries $\phi = 0$, with small and large values of $|\nabla \phi|$ leading to wide and narrow transition regions, respectively (Fig.~\ref{fig:LevelSet}a, right).

We propose to regularize the material density by forcing $|\nabla \phi|$ to be unity in a narrow band $\Omega_\mathrm{eik}$ of width $w$ along the material boundaries defined by $\phi = 0$. In this way, $\phi$ becomes a signed distance function to the material boundary in the narrow band, thereby constraining the gradient of $\rho$ to be constant along the interface. To ensure that the narrow band covers the near-entirety of the transition region where $\rho$ goes from 0 to 1, we choose $w = 10 \delta$ so that $\rho = \mathrm{sigmoid}(\pm w/2\delta) = \mathrm{sigmoid}(\pm 5) \simeq 0$ or $1$ along the edge of the narrow band. To illustrate the effect of such regularization, we consider the previous random instance of the neural network $\phi = \bar{\phi}(\mathbf{x},\boldsymbol{\theta}_\phi)$ and enforce the constraint $|\nabla \phi| = 1$ in the narrow band $\Omega_\mathrm{eik}$ along its zero level-set (Fig.~\ref{fig:LevelSet}b, left and right). The zero level-set is kept fixed to facilitate comparison with the unregularized case (Fig.~\ref{fig:LevelSet}a). With $\phi$ now behaving like a signed-distance function in the narrow band, a uniform transition thickness for $\rho$ along all material boundaries is achieved, without large regions of intermediate density values (Fig.~\ref{fig:LevelSet}b, center).

In practice, we implement this regularization into our PINN-based TO framework by including an `eikonal' loss term $\mathcal{L}_\mathrm{eik}$ in \eqref{eq:TotalLoss}, which takes the form
\begin{equation}
\mathcal{L}_\mathrm{eik}(\boldsymbol{\theta}_\phi) = \frac{1}{|\Omega_\mathrm{eik}^d|} \sum_{\mathbf{x}_i \in \Omega_\mathrm{eik}^d} \left(| \nabla \phi (\mathbf{x}_i)| - 1 \right)^2,
\label{eq:EikonalLoss}
\end{equation}
where $\Omega_\mathrm{eik}^d = \{\mathbf{x}_i \in \Omega^d : |\phi(\mathbf{x}_i)| < w/2\}$. The aim of this term is to penalize deviations away from the constraint $|\nabla \phi| = 1$ in the narrow band $\Omega_\mathrm{eik}$ of width $w$ along the interface defined by the zero level-set $\phi = 0$. Because finding the subset of collocation points $\mathbf{x}_i$ in $\Omega^d$ belonging to the true narrow band of width $w$ at every step of the training process would be too expensive, we instead relax the domain over which the constraint $|\nabla \phi| = 1$ is active by utilizing the subset $\Omega_\mathrm{eik}^d$ of collocation points that satisfy $|\phi(\mathbf{x}_i)| < w/2$. As the constraint $|\nabla \phi| = 1$ is progressively better satisfied during the training process, $\Omega_\mathrm{eik}^d$ will eventually overlap the true narrow band of width $w$ along the zero level-set of $\phi$ (Fig.~\ref{fig:LevelSet}c). 

We call this approach eikonal regularization, since the constraint $|\nabla \phi| = 1$ in the narrow band takes the form of an eikonal equation. \textcolor{black}{However, in contrast to the way the eikonal equation is employed in classical level-set methods \citep{adalsteinsson1995,osher2004} and recent works on neural signed distance functions \citep{gropp2020}, our eikonal regularization does not require $\phi$ to vanish on a specified boundary. That is, we do not enforce an explicit boundary condition $\phi = 0$ along a given curve (in 2D) or surface (in 3D), which is typically needed to solve the eikonal equation.} Rather, we let the zero level-set of $\phi$ freely evolve during the training process, eventually revealing the material boundaries delineating the hidden voids or inclusions. \textcolor{black}{To our knowledge, this is the first time that the eikonal equation is used without an explicit boundary condition.}

\subsection*{Setup of numerical experiments}
\label{sec:SetupNumericalExperiments}

We evaluate our TO framework on \textcolor{black}{numerous} challenging test cases \textcolor{black}{in 2D and 3D involving different types of physics (mechanical and thermal), several shapes and materials for the body $\mathcal{B}$, and various numbers and shapes of hidden voids and inclusions. These test cases are described in Materials and Methods, and listed in Tabs.~\ref{tab:ElasticMatrixCases}, \ref{tab:ElasticLayerCases}, \ref{tab:ThermalMatrixCases}. As a substitute for real experiments, we use the finite-element method (FEM) software Abaqus to compute the response of the body to the applied loading} and generate the measurement data for each case; \textcolor{black}{details are provided in the Materials and Methods}. Using this \textcolor{black}{synthetic} measurement data, we run our TO framework to discover the number, locations, and shapes of the hidden voids or rigid inclusions. We then compare the obtained results with the ground truth -- the voids or inclusions originally fed into Abaqus -- to assess the efficacy of our framework. \textcolor{black}{A comprehensive description of the implementation details and the training procedure for the TO framework is provided in the Supplementary Materials.}

\subsection*{\textcolor{black}{Prototypical examples}}
\label{sec:PrototypicalExamples}

\begin{figure*}
\centering
\includegraphics[width=\textwidth]{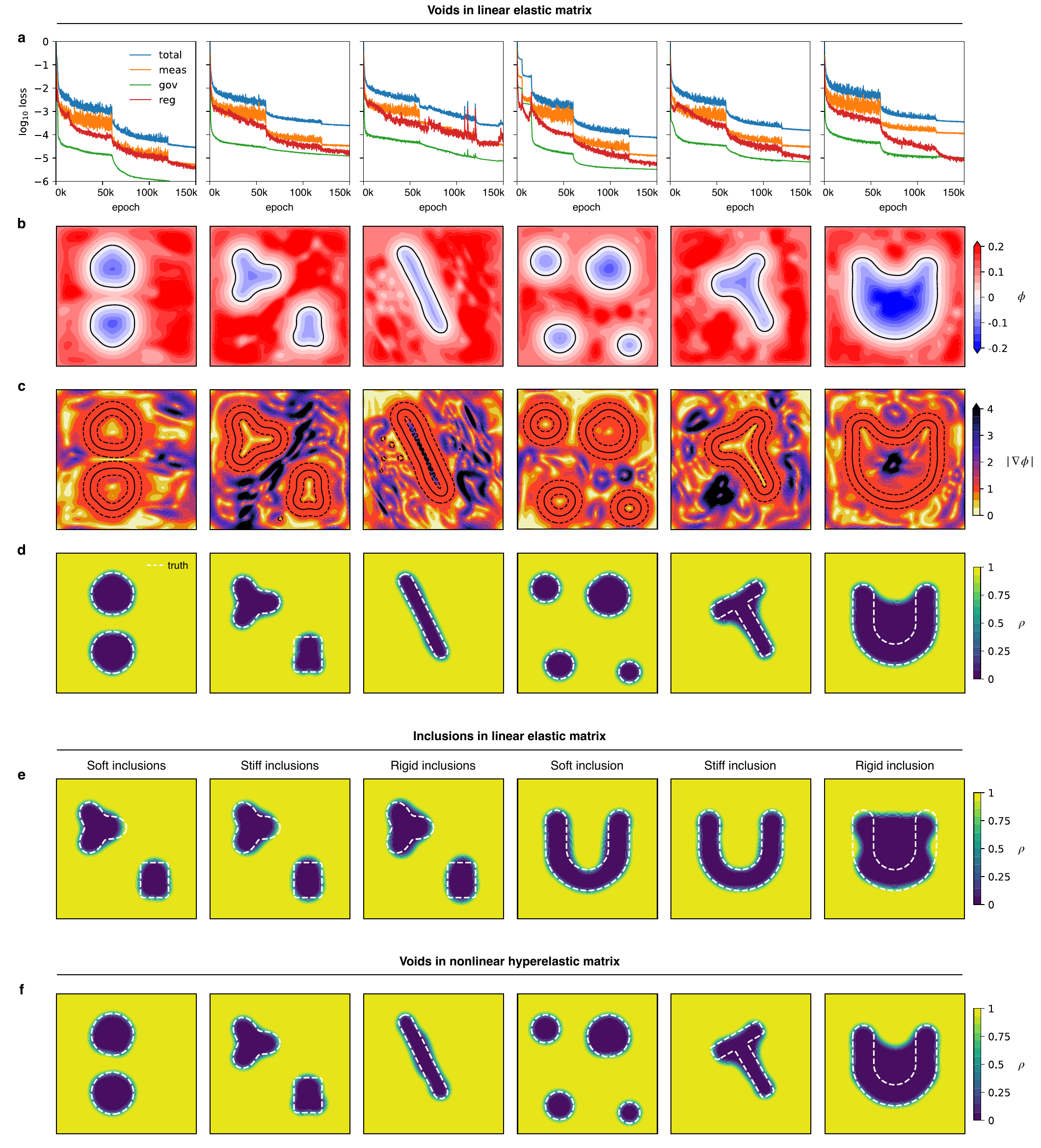}
\caption{\textbf{Identification of voids and inclusions in elastic matrices.} A linear elastic matrix containing voids (\textbf{a-d}): \textbf{a}, The various loss components that enforce the solution to match the surface measurement data, satisfy the governing equations, and obey the eikonal regularization, are being minimized during the training process. \textbf{b},\textbf{c}, The final level-set function $\phi$ and its gradient magnitude $|\nabla \phi|$ show the effect of the eikonal regularization, making $\phi$ a signed distance function in narrow band along the interface. \textbf{d}, The final material density $\rho$ reveals the number, locations, and shapes of the hidden voids, which are compared with the ground truth shown in dotted white lines. \textbf{e}, The final material density predictions in the case of a linear elastic matrix containing soft, stiff or rigid inclusions. \textbf{f}, The final material density predictions in the case of a nonlinear hyperelastic matrix containing voids subject to large stretches.}
\label{fig:NCS_Ela}
\end{figure*}

\textcolor{black}{\textbf{Voids in square elastic matrix.}} We first apply our framework to \textcolor{black}{the 2D square linear elastic matrix setup of Fig.~\ref{fig:Geometry}a in the presence of hidden voids} (cases 1, 2, 4, 9, 11, 16, 18 in Tab.~\ref{tab:ElasticMatrixCases}). As the various loss components are minimized during training (Fig.~\ref{fig:NCS_Ela}a), the material density $\rho$ evolves and splits in a way that progressively reveals the number, locations, and shapes of the hidden voids (Fig.~S2 and \href{https://www.dropbox.com/s/tm26buzgn0w35s1/Movie1.mp4?dl=0}{Movie 1} in Supplementary Materials), without advance knowledge of their topology. By the end of the training, the transition regions where the material density goes from 0 to 1 have uniform thickness along all internal boundaries (Fig.~\ref{fig:NCS_Ela}d), thanks to the eikonal regularization that encourages the level-set gradient $\nabla \phi$ to have unit norm in a band along the material boundaries $\phi = 0$ (Fig.~\ref{fig:NCS_Ela}b,c). The agreement between the final inferred shapes and the ground truth is remarkable, with our framework able to recover intricate details such as the three lobes and the concave surfaces of the star-shaped void (Fig.~\ref{fig:NCS_Ela}d, second from left), or the exact aspect ratio and location of a thin slit (Fig.~\ref{fig:NCS_Ela}d, third from left). The stress and strain fields of the deformed matrix are also obtained as a byproduct of the solution process (Fig.~S3 in Supplementary Materials).

The only case that is not completely identified is the U-shaped void (Fig.~\ref{fig:NCS_Ela}d, first from right), which demonstrates an important limit of the generic inverse problem. In order to be uniquely identifiable theoretically, each void must be simply-connected (see Supplementary Materials).  Voids that are close to having disconnected material in their interior, such as the U, cause the inverse problem to be ill-conditioned, and thus challenging for numerical algorithms to solve. In the U-shaped void, this can be seen by the minuscule influence the inner lobe of the U exerts on the outer surface displacements, due to its negligible level of strain and stress (Fig.~S10 in Supplementary Materials).

\textcolor{black}{To investigate the effect of measurement noise on the accuracy of the results, we repeat cases 4, 9, and 18 in the presence of white Gaussian measurement noise of standard deviation $\sigma_\mathrm{noise}$, demonstrating strong robustness of the results with respect to noise (Fig.~S6 in Supplementary Materials). We also analyze quantitatively the relationship between noise level and results accuracy, resulting in some practical guidelines to minimize the effect of noise (see Supplementary Materials).} Finally, our framework maintains accurate results when reducing the number of surface measurement points or restricting measurements to a few surfaces (Figs.~S7, S8, S9 in Supplementary Materials).

\textcolor{black}{\textbf{Inclusions in square elastic matrix.}} Next, we consider cases involving linear elastic and rigid inclusions in the linear elastic matrix (cases 5, 6, 7, 12, 13, 14 in Tab.~\ref{tab:ElasticMatrixCases}). Our framework successfully identifies the inclusions in almost all cases (Fig.~\ref{fig:NCS_Ela}e). Inferred displacements and stresses of the deformed matrix (Fig.~S4 in Supplementary Materials) confirm the intuition that voids or soft inclusions soften the matrix while stiff or rigid inclusions harden the matrix. The U-shaped soft and stiff elastic inclusions (Fig.~\ref{fig:NCS_Ela}e, second and third from right) are better detected by the framework than their void or rigid counterparts (Fig.~\ref{fig:NCS_Ela}d, first from right and Fig.~\ref{fig:NCS_Ela}e, first from right), since an elastic inclusion induces some strain and stress on the inner lobe (Fig.~S10 in Supplementary Materials).

\textcolor{black}{\textbf{Voids in square hyperelastic matrix.} We} consider cases with the same void shapes considered previously, but this time embedded in a soft, incompressible neo-Hookean hyperelastic matrix (cases 3, 8, 10, 15, 17, 19 in Tab.~\ref{tab:ElasticMatrixCases}). 
The geometries are identified equally well (Fig.~\ref{fig:NCS_Ela}f) in this large deformation regime (\textcolor{black}{see the large strains in} Fig.~S5 in Supplementary Materials) as with linear elastic materials, which illustrates the ability of the framework to cope with nonlinear governing equations without any added complexity in the formulation or the implementation.

\textcolor{black}{\textbf{Rigid substrate under elastic layer.}} We finally apply our framework to the periodic elastic layer setup of Fig.~\ref{fig:Geometry}b, where a linear elastic material covers a hidden rigid substrate (cases 23, 24, 25 in Tab.~\ref{tab:ElasticLayerCases}).
\begin{figure}
\centering
\includegraphics[width=0.49\textwidth]{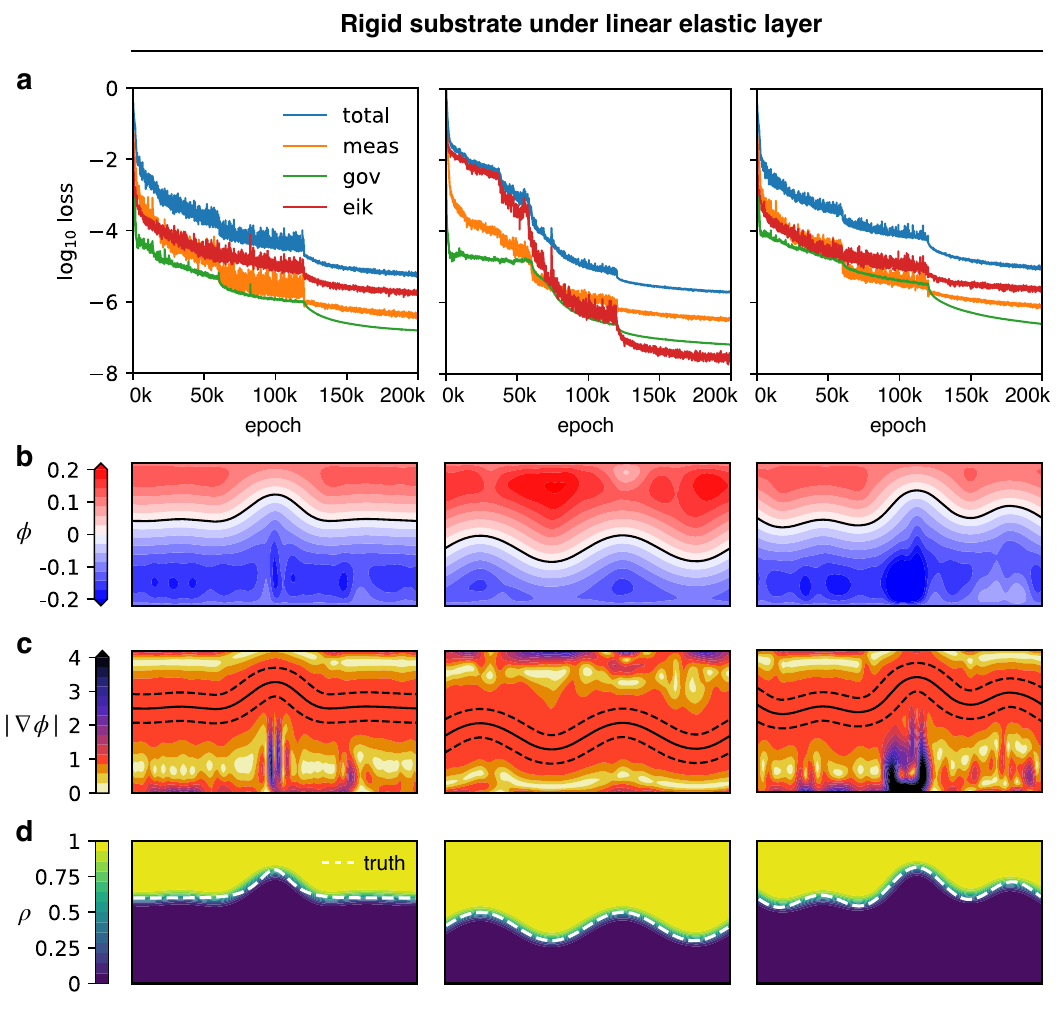}
\caption{\textbf{Identification of substrate shape underneath a periodic linear elastic layer.} \textbf{a}, The various loss components that enforce the solution to match the surface measurement data, satisfy the governing equations, and obey the eikonal regularization, are being minimized during the training process. \textbf{b},\textbf{c}, The final level-set function $\phi$ and its gradient magnitude $|\nabla \phi|$ show the effect of the eikonal regularization, which makes $\phi$ a signed distance function in narrow band along the material boundary. \textbf{d}, The final material density $\rho$ reveals the shape of the buried rigid substrate.}
\label{fig:NCS_PeriodicEla}
\end{figure}
Contrary to the matrix problem, this setup only provides access to measurements on the top surface, and the hidden geometry to be discovered is not completely surrounded by the elastic material. Our TO framework is nevertheless able to detect the correct depth and shape of the hidden substrate (Fig.~\ref{fig:NCS_PeriodicEla}), demonstrating its versatility.

\subsection*{\textcolor{black}{Advanced examples}}
\label{sec:AdvancedExamples}

\begin{figure*}
\centering
\includegraphics[width=\textwidth]{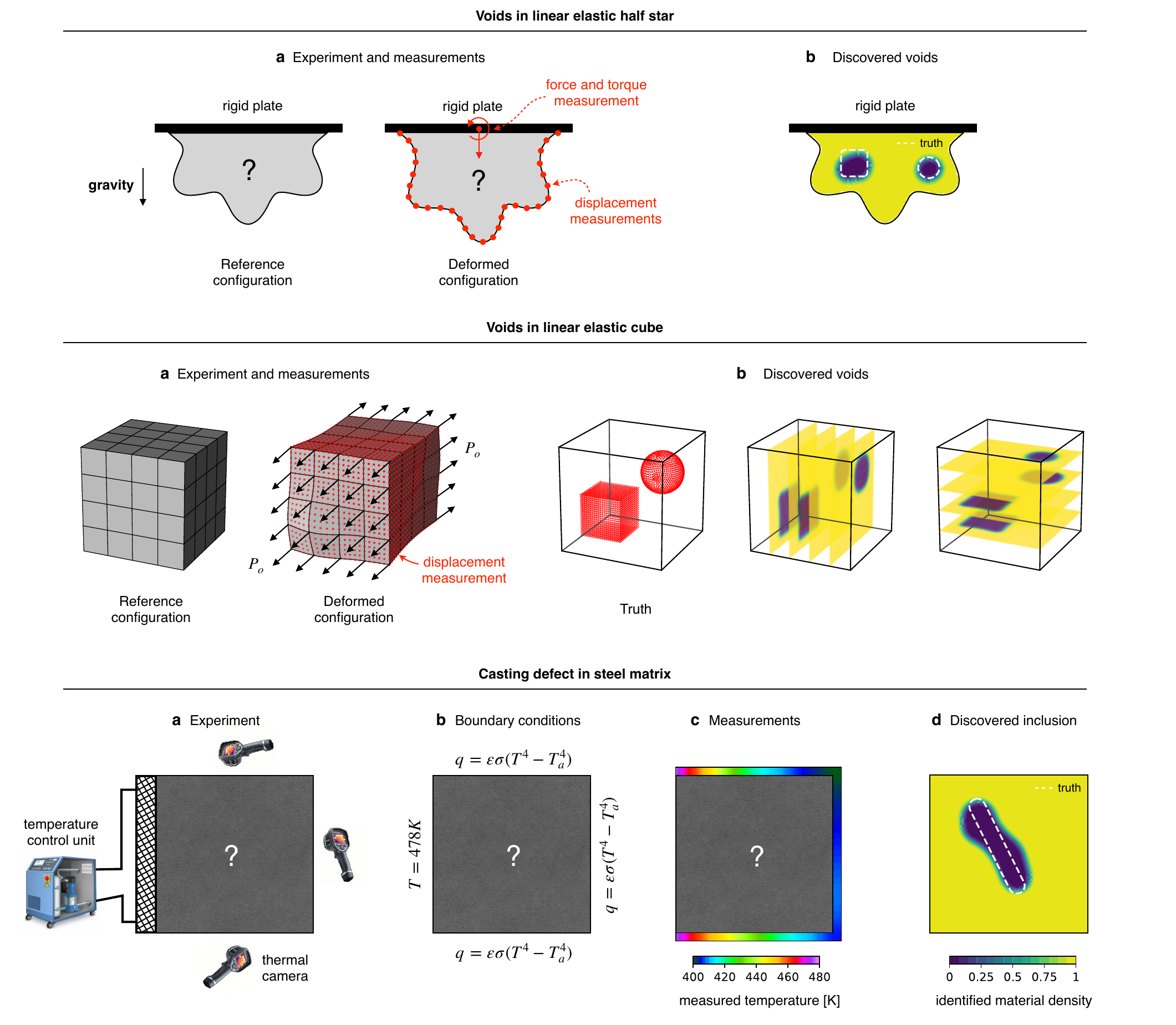}
\caption{\textcolor{black}{\textbf{Advanced examples. Top row: Identification of voids in a linear elastic half star.} \textbf{a}, Setup of the numerical experiment, where a half star is glued underneath a rigid plate and pulled by gravity. The measurements consist of the displacement along the wavy boundary as well as the resultant force and torque on the rigid plate. \textbf{b}, The final material density identified by our framework reveals the hidden voids. \textbf{Middle row: Identification of voids in a 3D linear elastic cube.} \textbf{a}, Setup of the numerical experiment, which is a 3D extension of the elastic square matrix example. The cube is pulled by a known uniform traction on two opposite sides, and the resulting displacement along the outer boundary is measured. \textbf{b}, The final material density identified by our framework reveals the hidden voids.} \textbf{\textcolor{black}{Bottom row:} Identification of a slender casting defect in a steel matrix.} \textbf{a}, Setup of the numerical experiment, which can be realized in practice with a temperature control unit and a thermal camera. \textbf{b}, The left side is heated to a constant temperature, modeled by a Dirichlet BC, while the remaining three sides are left exposed to air, modeled by a radiation BC. \textbf{c}, The measurements consist of the temperature profiles on the three sides exposed to air. \textbf{d}, The final material density identified by our framework reveals the slender void.}
\label{fig:Advanced}
\end{figure*}

\textcolor{black}{Through several additional examples, we demonstrate how the versatility of PINNs allows our approach to be easily tailored} to various scenarios involving \textcolor{black}{complex object shapes,} 3D geometries, partial information, \textcolor{black}{and} other physical loading types. \textcolor{black}{For example, our framework successfully identifies hidden voids in a linear elastic half-star-shaped object pulled by gravity rather than applied surface stresses (Fig.~\ref{fig:Advanced}, top row; case 21 in Tab.~\ref{tab:ElasticMatrixCases}). For this case, we derived a new approach to implement hard-constrained traction-free BCs along curved boundaries in PINNs (see Supplementary Materials). We also applied our method to a 3D linear elastic cube pulled by opposite surface tractions (Fig.~\ref{fig:Advanced}, middle row; case 22 in Tab.~\ref{tab:ElasticMatrixCases}), successfully identifying the hidden cubical and spherical voids within. Moreover,} we demonstrate the practical applicability of our framework on a realistic and easily implementable \textcolor{black}{thermal imaging} setup in which we detect a casting defect in a hypothetical steel matrix by simply heating one side and measuring the temperature distribution on the remaining sides exposed to air with a thermal camera (Fig.~\ref{fig:Advanced}\textcolor{black}{, bottom row; case 26 in Tab.~\ref{tab:ThermalMatrixCases}}). \textcolor{black}{Finally,} in a separate \textcolor{black}{thermal imaging} setup, we identify a hidden inclusion in a nonlinearly conducting matrix \textcolor{black}{without knowing the BC or measurements along one side} (\textcolor{black}{Fig.~S11 in Supplementary Materials; cases 27, 28 in Tab.~\ref{tab:ThermalMatrixCases}}), which reveals our method's ability to \textcolor{black}{identify hidden objects despite the complete lack of information along an entire side}.  The ability to uniquely construct the hidden objects from data on only part of the boundary is guaranteed by the Theorem stated in the Problem formulation section.

\section*{Discussion}

\begin{figure*}
\centering
\includegraphics[width=\textwidth]{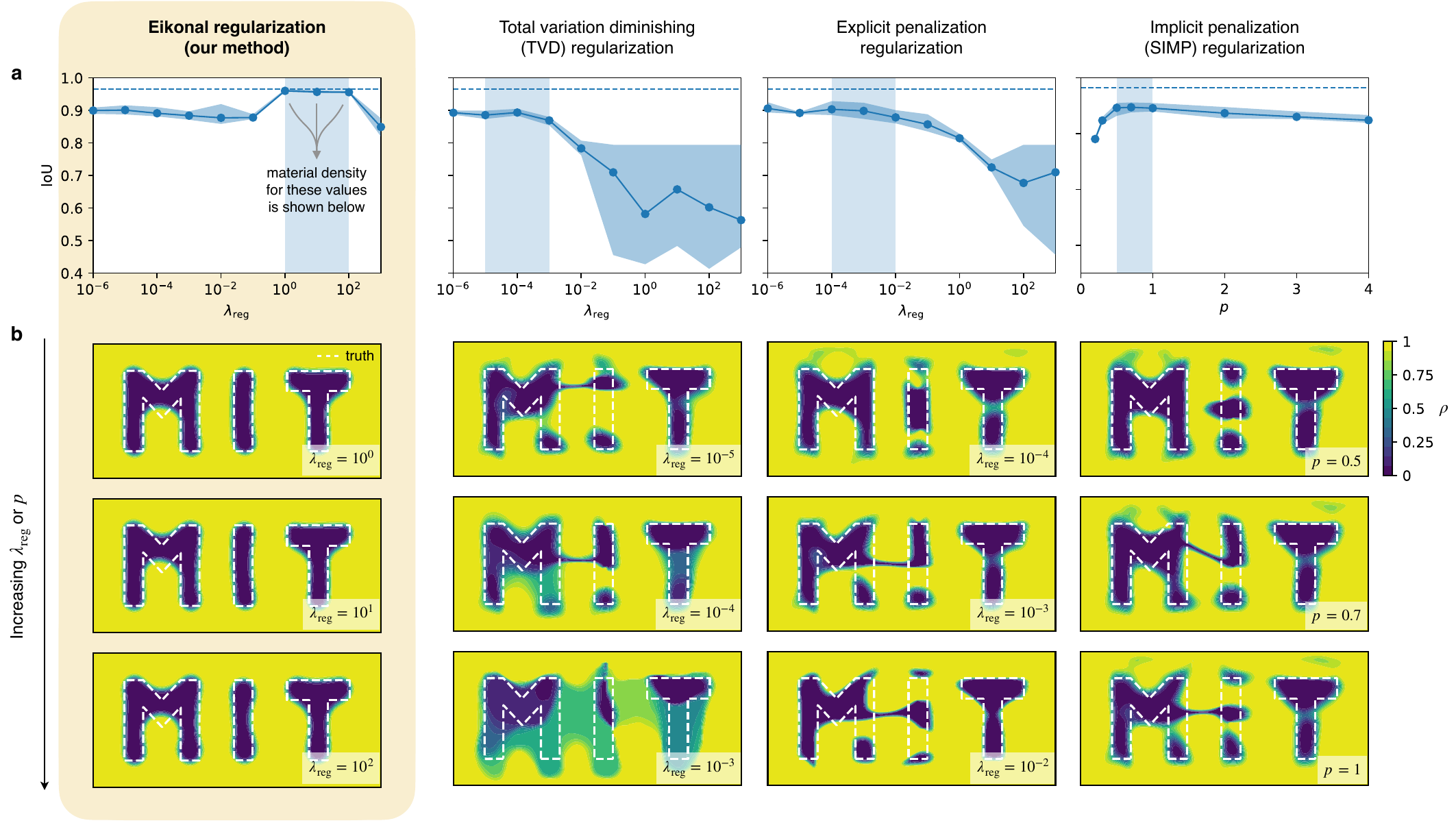}
\caption{\textbf{Comparison between eikonal regularization and alternative regularizations.} \textbf{a}, The eikonal regularization achieves a high IoU (intersection over union, a geometry detection accuracy metric equal to 1 in the perfect case) above 0.95 for any value of the regularization weight $\lambda_\mathrm{reg}$ within a range spanning three orders of magnitude. The results are consistent over 4 random initializations of the neural networks parameters, with the circles reporting the average value and the shade reporting the highest and lowest values. By contrast, the total variation diminishing (TVD), explicit penalization, and Solid Isotropic Material with Penalization (SIMP) regularizations never exceed IoU values above 0.93, with larger variability among realizations. \textbf{b}, The final material density $\rho$ obtained with each regularization mechanism for various values of the regularization weight $\lambda_\mathrm{reg}$ or exponent $p$ (shown in \textbf{a} by the shaded areas) demonstrates the efficacy of the eikonal regularization. The evolution of the solution during training using the eikonal regularization and $\lambda_\mathrm{reg} = 1$ is shown in \href{https://www.dropbox.com/s/ql1ko7e6kpf0cjr/Movie2.mp4?dl=0}{SI, Movie 2}.}
\label{fig:NCS_Reg}
\end{figure*}

Like any material density-based TO method, the success of our PINN-based framework hinges on the presence of an appropriate regularization mechanism to penalize intermediate density values. Although we have shown that our novel eikonal regularization leads to consistently accurate results, other regularization approaches have been employed in classical adjoint-based TO methods \citep{dorn2006,sigmund2013}. These include the total variation diminishing (TVD) regularization \citep{chan2004,mei2016} that penalizes the $L_1$ norm of the density gradient $\nabla \rho$, the explicit penalization regularization \citep{allaire1993} that penalizes the integral over the domain of $\rho (1-\rho)$, and the Solid Isotropic Material with Penalization (SIMP) approach \citep{bendsoe1989} that relates material properties such as the shear modulus and the material density through a power-law with exponent $p$. The latter is the most popular regularization mechanism in structural optimization \citep{bendsoe2003}. However, when implemented in our PINN-based framework for the detection of hidden geometries, these methods yield inferior results to the eikonal regularization (Fig.~\ref{fig:NCS_Reg}). Indeed, we compare all four approaches on a challenging test case involving a linear elastic rectangular matrix pulled from the top and bottom and containing soft inclusions in the shape of the letters M, I, and T (case 20, Tab.~\ref{tab:ElasticMatrixCases}). {The measurements consist of the displacement along the outer boundary, similar to the previous square matrix examples.} We consider different values of the regularization weight $\lambda_\mathrm{reg}$ (for the eikonal, TVD and explicit penalization regularizations) and the exponent $p$ (for the SIMP regularization), and solve the inverse problem using four random initializations of the neural networks in each case. Not only was the eikonal regularization the only one to find the right shapes, it did so over three orders of magnitude of $\lambda_\mathrm{reg}$ \textcolor{black}{and consistently over all four trials}, demonstrating a desirable robustness with respect to $\lambda_\mathrm{reg}$ (Fig.~\ref{fig:NCS_Reg}). Finally, we note that adjoint-based methods using TVD or other types of regularization limit themselves to simple shapes like squares and circles, or fail to find the right number of shapes \citep{ameur2004,mei2016,mei2021}.

\textcolor{black}{The TO method that we have introduced uses PINNs to solve the inverse problem. Nonetheless, our proposed eikonal regularization is potentially applicable to other imaging frameworks relying on a material density parametrization of the geometry, including those based on adjoint methods. While these methods may offer greater computational efficiency \cite{mowlavi2023,du2023}, the ease of implementation of PINNs is a significant advantage that greatly enhances the practical applicability of our method. Naturally, PINNs are not devoid of challenges such as hyperparameter selection or difficult enforcement of singular BCs; but these challenges are to some extent shared by all optimization frameworks, and improvements to PINNs are continuously proposed (for example, see \cite{cho2024}). Moreover, PINNs capitalize on the inherent robustness of overparameterized neural networks to sparse and noisy data \cite{li2020,clark2023,molnar2023}, enabling our PINN-based framework to discover hidden shapes accurately even in the presence of sparse or noisy measurements (Figs.~S6, S7, S8, S9 in Supplementary Materials).}

Finally, some imaging methods assume the unknown geometry is comprised of elementary shapes like circles and ellipses as opposed to a material density field. This yields a simpler search space with a few scalar parameters defining the locations and sizes of these predefined shapes \citep{schnur1992,mellings1995,zhang2022}. Notably, Zhang et al.~\cite{zhang2022} implemented this approach with PINNs in order to detect elliptical voids and inclusions in linear-elastic rectangular bodies. The general applicability of this family of methods is limited by the requirement that the number and types of shapes be given ahead of time. Such a priori knowledge is not required in our proposed TO framework. 


\textcolor{black}{A general limitation of the single-loading setup considered in this paper is that the material properties of the matrix and inclusions are assumed to be known ahead of time, which is a reasonable assumption in many applications. However, if such knowledge were unavailable, we expect from theoretical results on the Calder\'on problem that data from multiple loadings would enable the solver to simultaneously infer the material properties of the inclusions as well as their shapes, as realized for example in electrical impedance tomography \cite{cheney1999,adler2021}.}

In conclusion, we have presented a PINN-based TO framework with a novel eikonal regularization, which we have applied to the noninvasive detection of hidden inclusions. By representing the geometry through a material density field combined with the eikonal regularization, our framework is able to discover the number, shapes and locations of hidden structures, without any prior knowledge required regarding the number or the types of shapes to expect. Finally, \textcolor{black}{the introduction of the eikonal regularization opens a pathway for PINNs to be applied to a wide range of design optimization problems involving \textit{a priori} unknown topology and constrained by physical governing equations}. These include, for instance, the design of lenses that achieve targeted optical properties \citep{molesky2018,ma2021} or the design of structures and metamaterials that exhibit desirable mechanical, acoustic, or thermal properties \citep{bendsoe2003,zegard2016,kadic2019,kollmann2020,akerson2022,zhang2023fast,zhang2024chrono}.

\section*{Materials and methods}

\subsection*{List of experiments}

\textcolor{black}{To evaluate our TO framework, we study a wide range of test cases falling under three categories: an elastic matrix containing voids or inclusions, an elastic layer sitting on top of a rigid substrate, and a thermally conductive matrix containing a perfectly conductive or insulating inclusion. The precise mathematical formulation of the governing PDEs, applied BCs, and measurements corresponding to each case is provided in the Supplementary Materials.}

\textcolor{black}{\textbf{Elastic matrix.} The corresponding cases are listed in Tab.~\ref{tab:ElasticMatrixCases} and sorted according to the shape of the matrix, the number and shape of inclusions (to be discovered by the TO framework), and the constitutive properties of the matrix and inclusions. Cases 1 to 19 correspond to the 2D square matrix setup described in Fig.~\ref{fig:Geometry}a, where a traction is applied to the left and right sides of the matrix and displacement measurements are acquired along all four sides. Case 20 is a slight modification of that setup involving a rectangular matrix pulled from top and bottom. Case 21 is the setup described in the first row of Fig.~\ref{fig:Advanced}, where a half-star-shaped matrix is glued to a rigid plate and is pulled by gravity. The measurements consist of the displacement along the wavy boundary as well as the resultant force and torque on the rigid plate. Finally, case 22 is the setup described in the second row of Fig.~\ref{fig:Advanced}, where a 3D cube is subject to a traction on two opposite sides and displacement measurements are acquired along all six sides.}

\begin{table}
\centering
\setlength{\tabcolsep}{6pt}
\begin{tabular}{ l l l l l }
\toprule
Case & \multirow{2}{1cm}{Matrix shape} & \multirow{2}{1cm}{Inclusion shapes} & \multirow{2}{1cm}{Matrix material} & \multirow{2}{1cm}{Inclusion material} \\ \\
\toprule
1* & Square & \multirow{1}{2cm}{One circle} & LE & V \\
\midrule
2 & \multirow{2}{*}{Square} & \multirow{2}{2cm}{Two circles} & LE & V \\
3 & & & HE & V \\
\midrule
4*$^+$ & \multirow{5}{*}{Square} & \multirow{5}{2cm}{One star, one rectangle}  & LE & V \\
5 & & & LE & LE-soft \\
6 & & & LE & LE-stiff \\
7 & & & LE & R \\
8 & & & HE & V \\
\midrule
9*$^+$ & \multirow{2}{*}{Square} & \multirow{2}{2cm}{One slit} & LE & V \\
10 & & & HE & V \\
\midrule
11 & \multirow{5}{*}{Square} & \multirow{5}{2cm}{One U} & LE & V \\
12 & & & LE & LE-soft \\
13 & & & LE & LE-stiff \\
14 & & & LE & R \\
15 & & & HE & V \\
\midrule
16 & \multirow{2}{*}{Square} & \multirow{2}{2cm}{One T} & LE & V \\
17 & & & HE & V \\
\midrule
18$^+$ & \multirow{2}{*}{Square} & \multirow{2}{2cm}{Four circles} & LE & V \\
19 & & & HE & V \\
\midrule
\multirow{2}{*}{20} & \multirow{2}{*}{Rectangle} & \multirow{2}{2cm}{One M, one I and one T} &  \multirow{2}{*}{LE} & \multirow{2}{*}{LE-soft} \\ 
\\
\midrule
\multirow{2}{*}{21} & \multirow{2}{*}{Half-star} & \multirow{2}{2cm}{One circle, one rectangle} &  \multirow{2}{*}{LE} & \multirow{2}{*}{V} \\
\\
\midrule
\multirow{2}{*}{22} & \multirow{2}{*}{Cube (3D)} & \multirow{2}{2cm}{One sphere, one cube} &  \multirow{2}{*}{LE} & \multirow{2}{*}{V} \\
\\
\bottomrule
\end{tabular}
\caption{List of all elastic matrix cases, classified according to the geometry and elastic properties of the matrix and inclusions. LE: linear elastic with $E = E_0$ and $\nu = 0.3$; HE: incompressible neo-Hookoean hyperelastic with $\mu = 0.38$; LE-soft: linear elastic with $E = E_0/5$, $\nu = 0.3$; LE-stiff: linear elastic with $E = 5E_0$, $\nu = 0.3$; R: rigid; V: void. The star * and the plus $^+$ indicate that the corresponding case was evaluated using a varying number of surface measurement points and a non-zero amount of measurement noise, respectively. \label{tab:ElasticMatrixCases}}
\end{table}

\textcolor{black}{\textbf{Elastic layer.} The corresponding cases are listed in Tab.~\ref{tab:ElasticLayerCases} and sorted according to the shape of the bottom substrate (to be discovered by the TO framework). The general setup is shown Fig.~\ref{fig:Geometry}b, where the rigid substrate is hidden underneath a layer of elastic material in a periodic domain. A normal pressure is applied to the top of the layer and the resulting displacement of the top surface is measured. This case simulates the detection of large rigid structures buried underneath a compliant material, a situation encountered in diverse fields such as archaeology or medicine.}

\begin{table}
\setlength{\tabcolsep}{6pt}
\begin{tabular}{ l l l l }
\toprule
Case & Geometry & Layer & Substrate \\ 
\toprule
23 & Sinusoidal & LE & R \\
\midrule
24 & Pulse & LE & R \\
\midrule
25 & Random wave & LE & R \\
\bottomrule
\end{tabular}
\caption{List of all elastic layer cases, classified according to the shape of the substrate. LE: linear elastic with $E = E_0$ and $\nu = 0.3$; R: rigid. \label{tab:ElasticLayerCases}}
\end{table}

\textcolor{black}{\textbf{Thermally conductive matrix.} The corresponding cases are listed in Tab.~\ref{tab:ThermalMatrixCases} and sorted according to the type of thermal loading. Case 26 is the setup shown in the last row of Fig.~\ref{fig:Advanced}, where a square matrix with constant conductivity is prescribed a temperature on one side, while the remaining three sides are left exposed to air and their temperature is measured. This case aims to simulate the detection of a slender casting defect \cite{campbell2015} in a steel plate, using a heat transfer setup that is easily realisable experimentally with a temperature control unit and a thermal camera. The casting defect is modeled as a perfect insulator while the three exposed sides are modeled using a radiation BC, which takes the form of a nonlinear Robin BC. Cases 27 and 28 are the setup shown in Fig.~S11a of the Supplementary Materials, where a square matrix with temperature-dependent conductivity is assigned different temperatures on the left and right sides, and the top and bottom sides are insulated. It is assumed that one side is completely inaccessible in the sense that the applied BC and measurements are both unavailable, while temperature and flux measurements are acquired on the remaining three sides.}

\begin{table}
\centering
\setlength{\tabcolsep}{6pt}
\begin{tabular}{ l l l l l l }
\toprule
Case & \multirow{2}{1cm}{Loading type} & \multirow{2}{1cm}{Matrix shape} & \multirow{2}{1cm}{Inclusion shapes} & \multirow{2}{1cm}{Matrix material} & \multirow{2}{1cm}{Inclusion material} \\ \\
\toprule
26 & A & Square & One slit & C & PI \\
\midrule
27* & B & \multirow{2}{*}{Square} & \multirow{2}{*}{One slit}  & TD & PI \\
28* & B & & & TD & PC \\
\bottomrule
\end{tabular}
\caption{List of all thermal matrix cases, classified according to the loading type and thermal conductivity properties of the matrix and inclusions. A: Applied temperature difference between two opposite sides, remaining two sides are insulated. B: Applied temperature on one side, remaining three sides are left exposed to radiate heat. C: constant thermal conductivity; TD: temperature-dependent thermal conductivity; PI: perfectly insulating inclusion; PC: perfectly conductive inclusion. The star * indicates that the problem was solved assuming one of the four sides had unknown loading and no measurements. \label{tab:ThermalMatrixCases}}
\end{table}

\subsection*{FEM simulations}

\textcolor{black}{To obtain the measurement data for all cases considered in this paper, we perform FEM simulations in the software Abaqus, using its Standard (implicit) solver. Every 2D case is meshed using a linear density of 200 elements per unit length along each boundary, corresponding to between 25k to 80k total elements depending on domain size as well as number and shapes of voids or inclusions. For the mechanical loading experiments (Tabs.~\ref{tab:ElasticMatrixCases} and \ref{tab:ElasticLayerCases}), we employ bilinear quadrilateral CPE4 plain-strain elements for the cases involving a linear elastic material, and their hybrid constant-pressure counterpart CPE4H for the cases involving a hyperelastic material. For the thermal loading experiments (Tab.~\ref{tab:ThermalMatrixCases}), we employ biquadratic DC2D8 diffusive heat transfer elements.}

\subsection*{Acknowledgements}

The authors acknowledge the MIT SuperCloud and Lincoln Laboratory Supercomputing Center for providing computing resources.

\bibliography{bibliography}

\end{document}


\title{Supplementary information for: \\ Detecting hidden structures from a static loading experiment: \\ topology optimization meets physics-informed neural networks}

\author{Saviz Mowlavi}
 \email{mowlavi@merl.com}
\affiliation{Mitsubishi Electric Research Laboratories, Cambridge, MA 02139, USA}
\affiliation{Department of Mechanical Engineering, MIT, Cambridge, MA 02139, USA}
\author{Ken Kamrin}%
 \email{kkamrin@mit.edu}
\affiliation{Department of Mechanical Engineering, MIT, Cambridge, MA 02139, USA}

\begin{abstract}
\end{abstract}

\maketitle

\section{Uniqueness of solutions}

We sketch a proof for the uniqueness of solutions to the elasticity imaging problem considered in this paper. For the specific case of a 2D linear elastic material with a single void, it has been proved that there exists at most one cavity which yields the same surface displacements and stresses on a finite portion of the external boundary \citep{ang1999}. Our approach is different and applicable to any 2D or 3D problem governed by analytic and elliptic PDEs, be they linear or nonlinear, heat conduction, elasticity, etc. Although we present the proof for the case of a single void or inclusion, the same reasoning readily generalizes to any number of voids or inclusions.

\subsection{Useful lemmas}

Before sketching the proof, we state a few lemmas that will be useful:
\begin{itemize}
\item \textbf{Lemma 1 (Cauchy–Kowalevski theorem)} Consider the Cauchy problem defined by a PDE in a connected domain $\Omega$ with boundary conditions specified by Cauchy data on a hypersurface $S \in \Omega$. (The Cauchy data defines the values of the solution and its normal derivatives of order up to $k-1$ on $S$, where $k$ is the order of the PDE.) If the functions defining the PDE, the Cauchy data, and the hypersurface are analytic, then the Cauchy problem has a unique analytic solution in a neighborhood of $S$ \citep{folland2020}.
\item \textbf{Lemma 2}. Any $C^2$ solution of a linear or nonlinear elliptic PDE is a real analytic function \citep{morrey1958}.
\item \textbf{Lemma 3 (identity theorem)}. Any two real analytic functions on a path-connected domain $\Omega$ that are equal on a finite and connected subset $\mathcal{S} \subset \Omega$, are necessarily equal in all of $\Omega$ \citep{krantz2002}. As a particular case, if a real analytic function defined on $\Omega$ vanishes in a subset $\mathcal{S} \subset \Omega$, then it necessarily vanishes in all of $\Omega$.
\end{itemize}

\subsection{Setup}

\begin{figure}[tb]
\centering
\includegraphics[width=\textwidth]{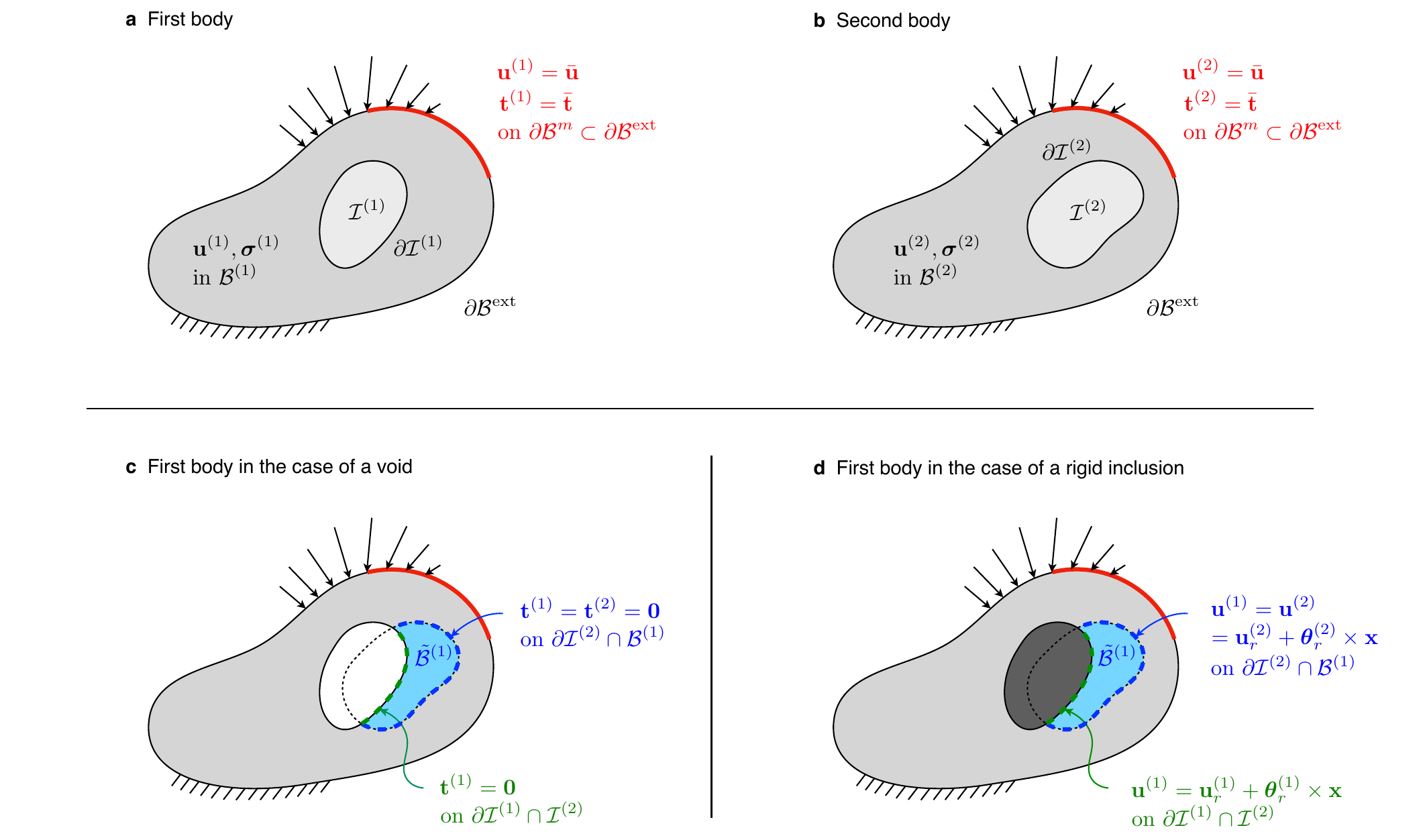}
\caption{\textbf{Setup for the proof of the uniqueness of solutions.} \textbf{a},\textbf{b}, Two identically-shaped bodies with different inclusions are assumed to have the same surface displacements and tractions along a portion $\partial \mathcal{B}^m$ of their outer boundary. \textbf{c}, In the case of a void, the blue region $\partial \tilde{\mathcal{B}}^{(1)}$ of the first body has vanishing stress. \textbf{d}, In the case of a rigid inclusion, the blue region $\partial \tilde{\mathcal{B}}^{(1)}$ of the first body undergoes a rigid displacement.}
\label{fig:Proof}
\end{figure}

Consider two bodies $\mathcal{B}^{(1)}$ and $\mathcal{B}^{(2)}$ with the same material properties and sharing the same external boundary $\partial \mathcal{B}^\mathrm{ext}$ (Supplementary Fig.~\ref{fig:Proof}a,b). Each body contains a single smooth void or rigid inclusion, characterized by the connected domains $\mathcal{I}^{(1)}$ and $\mathcal{I}^{(2)}$ with boundaries $\partial \mathcal{I}^{(1)}$ and $\partial \mathcal{I}^{(2)}$, respectively. 
Assume that there exists a finite connected segment $\partial \mathcal{B}^m \subset \partial \mathcal{B}^\mathrm{ext}$ on which the two bodies have identical surface displacements and tractions, yielding the Cauchy data $\mathbf{u}^{(1)} = \mathbf{u}^{(2)} = \bar{\mathbf{u}}$ and $\mathbf{t}^{(1)} = \mathbf{t}^{(2)} = \bar{\mathbf{t}}$ on $\partial \mathcal{B}^m$. In practice, such Cauchy data is obtained by knowing the applied displacement or traction and measuring the other quantity. Further, assume that the displacement and stress fields within each body, denoted by $\mathbf{u}^{(1)}$, $\boldsymbol{\sigma}^{(1)}$ and $\mathbf{u}^{(2)}$, $\boldsymbol{\sigma}^{(2)}$, are $C^2$ over $\mathcal{B}^{(1)}$ and $\mathcal{B}^{(2)}$, respectively. Finally, assume that $\bar{\mathbf{u}}$ and $\bar{\mathbf{t}}$ are analytic functions and $\bar{\mathbf{t}}$ does not vanish everywhere on $\partial \mathcal{B}^m$. We will prove that there cannot be two distinct shapes $\mathcal{I}^{(1)}$ and $\mathcal{I}^{(2)}$ yielding the same displacement and tractions on $\partial \mathcal{B}^m$. 

\subsection{Sketch of the proof}
The displacement fields $\mathbf{u}^{(1)}$ and $\mathbf{u}^{(2)}$ solve the equilibrium equations of elasticity. These equations are elliptic, and they are also analytic as defined in the sense of Lemma 1. As a result, Lemma 2 stipulates that $\mathbf{u}^{(1)}$ and $\mathbf{u}^{(2)}$ are real analytic functions over $\mathcal{B}^{(1)}$ and $\mathcal{B}^{(2)}$, respectively, and in addition, from Lemma 1, they must be equal in a neighborhood of $\partial \mathcal{B}^m$. Then, Lemma 3 implies that $\mathbf{u}^{(1)} = \mathbf{u}^{(2)}$, and therefore $\boldsymbol{\sigma}^{(1)} = \boldsymbol{\sigma}^{(2)}$, over $\mathcal{B}^{(1)} \cap \mathcal{B}^{(2)}$. We now focus on the behavior of the solution in body $\mathcal{B}^{(1)}$ and treat separately the cases of a void and a rigid inclusion. 

\subsubsection{Void}

Consider that each body contains a void, and let us focus on body $\mathcal{B}^{(1)}$. We will analyze the region $\tilde{\mathcal{B}}^{(1)} = \mathcal{B}^{(1)} \cap \mathcal{I}^{(2)}$ pictured in blue in Supplementary Fig.~\ref{fig:Proof}c. Assume first that the voids $\mathcal{I}^{(1)}$ and $\mathcal{I}^{(2)}$ overlap; then, the boundary delineating $\tilde{\mathcal{B}}^{(1)}$ contains one segment $\partial \mathcal{I}^{(1)} \cap \mathcal{I}^{(2)}$ and another segment $\partial \mathcal{I}^{(2)} \cap \mathcal{B}^{(1)}$. The traction $\mathbf{t}^{(1)}$ trivially vanishes on $\partial \mathcal{I}^{(1)} \cap \mathcal{I}^{(2)}$. Since $\boldsymbol{\sigma}^{(1)} = \boldsymbol{\sigma}^{(2)}$ in $\mathcal{B}^{(1)} \cap \mathcal{B}^{(2)}$, $\mathbf{t}^{(1)}$ also vanishes on $\partial \mathcal{I}^{(2)} \cap \mathcal{B}^{(1)}$. Assume now that the voids $\mathcal{I}^{(1)}$ and $\mathcal{I}^{(2)}$ do not overlap; then, the entire boundary delineating $\tilde{\mathcal{B}}^{(1)}$ is equal to $\partial \mathcal{I}^{(2)} \cap \mathcal{B}^{(1)}$, along which we just showed that $\mathbf{t}^{(1)}$ vanishes. Therefore, in both cases, the region $\tilde{\mathcal{B}}^{(1)}$ has zero traction along its entire boundary, implying that $\boldsymbol{\sigma}^{(1)}$ vanishes in $\tilde{\mathcal{B}}^{(1)}$.

Then, Lemma 3 implies that $\boldsymbol{\sigma}^{(1)}$ must vanish not only in $\tilde{\mathcal{B}}^{(1)}$, but also in the entire body $\mathcal{B}^{(1)}$.  This violates the fact that the surface traction $\mathbf{t}^{(1)}$ cannot vanish everywhere on $\partial \mathcal{B}^m$, so $\mathcal{I}^{(1)}$ and $\mathcal{I}^{(2)}$ must be identical. 

Note that to use Lemma 3 in the prior paragraph, we must assume that a path-connection exists between all points of $\tilde{\mathcal{B}}^{(1)}$ and any point in $\mathcal{B}^{(1)}$.  This occurs if each void is a simply connected domain.  As a counter-example, suppose a void contains a disconnected island of material within it (a ``rattler'' so to speak). We should not expect to be able to determine the shape of the rattler based on data from $\partial\mathcal{B}^{ext}$ because the rattler is completely disconnected from the rest of the body.  The path-connectedness caveat of Lemma 3 removes these cases from the proof. That is, only those void boundaries that are path-connected to $\partial\mathcal{B}^{ext}$ can be  uniquely identified from data on $\partial\mathcal{B}^{ext}$.

\subsubsection{Rigid inclusion}

Consider that each body contains a rigid inclusion, and let us focus on body $\mathcal{B}^{(1)}$. We will analyze the region $\tilde{\mathcal{B}}^{(1)} = \mathcal{B}^{(1)} \cap \mathcal{I}^{(2)}$ pictured in blue in Supplementary Fig.~\ref{fig:Proof}d. Assume first that the voids $\mathcal{I}^{(1)}$ and $\mathcal{I}^{(2)}$ overlap; then, the boundary delineating $\tilde{\mathcal{B}}^{(1)}$ contains one segment $\partial \mathcal{I}^{(1)} \cap \mathcal{I}^{(2)}$ and another segment $\partial \mathcal{I}^{(2)} \cap \mathcal{B}^{(1)}$. The displacement $\mathbf{u}^{(1)}$ on $\partial \mathcal{I}^{(1)} \cap \mathcal{I}^{(2)}$ corresponds to that of a rigid motion, i.e.~$\mathbf{u}^{(1)} = \mathbf{u}_r^{(1)} + \boldsymbol{\theta}_r^{(1)} \times \mathbf{x}$ for some fixed $\mathbf{u}_r^{(1)}$ and $\boldsymbol{\theta}_r^{(1)}$. Since $\mathbf{u}^{(1)} = \mathbf{u}^{(2)}$ in $\mathcal{B}^{(1)} \cap \mathcal{B}^{(2)}$, we also know that $\mathbf{u}^{(1)} = \mathbf{u}^{(2)} = \mathbf{u}_r^{(2)} + \boldsymbol{\theta}_r^{(2)} \times \mathbf{x}$ on $\partial \mathcal{I}^{(2)} \cap \mathcal{B}^{(1)}$ for some other fixed $\mathbf{u}_r^{(2)}$ and $\boldsymbol{\theta}_r^{(2)}$. However, given that these two rigid motions must coincide at $\partial \mathcal{I}^{(1)} \cap \partial \mathcal{I}^{(2)}$, we must have $\mathbf{u}_r^{(1)} = \mathbf{u}_r^{(2)}$ and $\boldsymbol{\theta}_r^{(1)} = \boldsymbol{\theta}_r^{(2)}$. Assume now that the voids $\mathcal{I}^{(1)}$ and $\mathcal{I}^{(2)}$ do not overlap; then, the entire boundary delineating $\tilde{\mathcal{B}}^{(1)}$ is equal to $\partial \mathcal{I}^{(2)} \cap \mathcal{B}^{(1)}$, along which we just showed that $\mathbf{u}^{(1)} = \mathbf{u}_r^{(2)} + \boldsymbol{\theta}_r^{(2)} \times \mathbf{x}$ on $\partial \mathcal{I}^{(2)} \cap \mathcal{B}^{(1)}$. Therefore, in both cases, the entire boundary of the region $\tilde{\mathcal{B}}^{(1)}$ undergoes a single rigid motion, implying that $\boldsymbol{\sigma}^{(1)}$ vanishes in $\tilde{\mathcal{B}}^{(1)}$. 

As before, Lemma 3 then implies that $\boldsymbol{\sigma}^{(1)}$ must vanish not only in $\tilde{\mathcal{B}}^{(1)}$, but also in the entire body $\mathcal{B}^{(1)}$. This violates the fact that the surface traction $\mathbf{t}^{(1)}$ cannot vanish everywhere on $\partial \mathcal{B}^\mathrm{ext}$, so $\mathcal{I}^{(1)}$ and $\mathcal{I}^{(2)}$ must be identical.  Note that, as in the case of voids, we require path-connectedness between $\tilde{\mathcal{B}}^{(1)}$ and any point in $\mathcal{B}^{(1)}$, which is assured by the assumption that the rigid inclusions are simply connected, i.e. have fully-rigid interiors.

\section{Governing equations}

In this Appendix, we describe the governing equations, boundary conditions, and measurement data for all cases studied in this paper. Cases listed in Tabs.~I and II and involving a linear elastic matrix are described in Section \ref{app:SmallDeformationLinearElasticity}, while those involving a hyperelastic matrix are described in Section \ref{app:LargeDeformationNonlinearHyperelasticity}. Cases listed in Tab.~III are described in Section \ref{app:ThermalExperiments}.

\subsection{Mechanical experiments}
\label{app:MechanicalExperiments}

\subsubsection{Small-deformation linear elasticity} 
\label{app:SmallDeformationLinearElasticity}

\textbf{Physical quantities.} We first consider the case where the elastic body and inclusions consist of linear elastic materials, with Young's modulus $E$ and Poisson's ratio $\nu$ for the body, and Young's modulus $\bar{E}$ and Poisson's ratio $\bar{\nu}$ for the inclusions. Voids and rigid inclusions correspond to the limits $\bar{E} \rightarrow 0$ and $\bar{E} \rightarrow \infty$, respectively. The deformation of the elastic body is described by the vector field $\boldsymbol{\psi}(\mathbf{x}) = (\mathbf{u}(\mathbf{x}), \boldsymbol{\sigma}(\mathbf{x}))$, where $\mathbf{u}(\mathbf{x})$ is the planar displacement field with components $u_i(\mathbf{x})$ and $\boldsymbol{\sigma}(\mathbf{x})$ is the Cauchy stress tensor with components $\sigma_{ij}(\mathbf{x})$. Indices $i$ and $j$ will hereafter range from 1 to 2 for 2D cases, and from 1 to 3 for 3D cases.

\textbf{Governing PDEs.} The governing PDEs comprise the equilibrium equations
\begin{equation}
\sum_{j}\frac{\partial \sigma_{ij}}{\partial x_j} = 0, \quad \mathbf{x} \in \Omega,
\label{eq:LinearEquilibrium}
\end{equation}
as well as a linear elastic constitutive law $F(\boldsymbol{\sigma}, \nabla \mathbf{u}, \rho) = 0$ that we will express in two different but equivalent ways, depending on whether the inclusions are softer or stiffer than the matrix. For voids and soft inclusions, we consider the constitutive law in stress-strain form,
\begin{equation}
\boldsymbol{\sigma} = \rho \left[ \lambda \, \mathrm{tr} (\boldsymbol{\epsilon}) \, \mathbf{I} + 2 \mu \, \boldsymbol{\epsilon} \right] + (1 - \rho) \left[ \bar{\lambda} \, \mathrm{tr} (\boldsymbol{\epsilon}) \, \mathbf{I} + 2 \bar{\mu} \, \boldsymbol{\epsilon} \right], \quad \mathbf{x} \in \Omega,
\label{eq:LinearStressStrainVoid}
\end{equation}
where $\boldsymbol{\epsilon} = (\nabla \mathbf{u} + \nabla \mathbf{u}^T)/2$ is the infinitesimal strain tensor, $\mathrm{tr}(\cdot)$ denotes the trace, $\lambda = E \nu/(1+\nu)(1-2\nu)$ and $\mu = E/2(1+\nu)$ are the Lam\'e constants of the body, and $\bar{\lambda} = \bar{E} \bar{\nu}/(1+\bar{\nu})(1-2\bar{\nu})$ and $\bar{\mu} = \bar{E}/2(1+\bar{\nu})$ are the Lam\'e constants of the inclusions. Notice that in the case of voids, the stress vanishes in the $\rho = 0$ regions. For stiff and rigid inclusions, 
we consider the constitutive law in the inverted strain-stress form, which differs in 2D and 3D due to the plane strain assumption. In 2D,
\begin{equation}
\boldsymbol{\epsilon} = \rho \left[ \frac{1+\nu}{E} \boldsymbol{\sigma} - \frac{\nu(1+\nu)}{E} \, \mathrm{tr} (\boldsymbol{\sigma}) \, \mathbf{I} \right] + (1-\rho) \left[ \frac{1+\bar{\nu}}{\bar{E}} \boldsymbol{\sigma} - \frac{\bar{\nu}(1+\bar{\nu})}{\bar{E}} \, \mathrm{tr} (\boldsymbol{\sigma}) \, \mathbf{I} \right], \quad \mathbf{x} \in \Omega,
\label{eq:LinearStressStrainInclusion2D}
\end{equation}while in 3D,
\begin{equation}
\boldsymbol{\epsilon} = \rho \left[ \frac{1+\nu}{E} \boldsymbol{\sigma} - \frac{\nu}{E} \, \mathrm{tr} (\boldsymbol{\sigma}) \, \mathbf{I} \right] + (1-\rho) \left[ \frac{1+\bar{\nu}}{\bar{E}} \boldsymbol{\sigma} - \frac{\bar{\nu}}{\bar{E}} \, \mathrm{tr} (\boldsymbol{\sigma}) \, \mathbf{I} \right], \quad \mathbf{x} \in \Omega.
\label{eq:LinearStressStrainInclusion3D}
\end{equation}
Notice that in the case of rigid inclusions, the strain vanishes in the $\rho = 0$ regions.

\textbf{Boundary conditions.} For the 2D square elastic matrix (Fig.~1a), the domain is $\Omega = [-0.5,0.5] \times [-0.5,0.5]$. The boundary conditions are
\begin{subequations}
\begin{alignat}{2}
\boldsymbol{\sigma}(\mathbf{x}) \mathbf{n}(\mathbf{x}) &= \pm P_o \mathbf{e}_1, &\quad &\mathbf{x} \in \{ \pm 0.5 \} \times [-0.5,0.5], \\
\boldsymbol{\sigma}(\mathbf{x}) \mathbf{n}(\mathbf{x}) &= \mathbf{0}, \quad &&\mathbf{x} \in [-0.5,0.5] \times \{ \pm 0.5 \},
\end{alignat} \label{eq:2DMatrixBCs}%
\end{subequations}
where $P_o/E = 0.01$, and the displacement $\mathbf{u}(\mathbf{x})$ is measured at locations $\partial \Omega^m$ distributed along the entire external boundary $\partial \Omega$. In the case of the M, I, T inclusions, the domain is $\Omega = [-1,1] \times [-0.5,0.5]$ and the boundary conditions \eqref{eq:2DMatrixBCs} are changed to account for the fact that the matrix is pulled from the top and bottom boundaries.

For the 2D half-star elastic matrix (Fig.~6, top row), the domain $\Omega$ is delineated by the top boundary $\partial\Omega_\mathrm{top} = \{ \mathbf{x} = (x,y) : x \in [-r^*, r^*], y = 0 \}$ and the wavy boundary $\partial\Omega_\mathrm{wavy} = \{ \mathbf{x} = (x,y) : x = r(\theta) \cos \theta, y = r(\theta) \sin \theta, r(\theta) = (r^* + a^* \sin 7 \theta), \theta \in [\pi, 2 \pi] \}$, with $r^* = 0.45$ and $a^* = 0.07$. Because the half-star is pulled by gravity, a body force $-g \mathbf{e}_2$ is added to the right-hand side of the equilibrium relation \eqref{eq:LinearEquilibrium}. The boundary conditions are
\begin{subequations}
\begin{alignat}{2}
\mathbf{u}(\mathbf{x}) &= \mathbf{0}, &\quad &\mathbf{x} \in \partial\Omega_\mathrm{top}, \\
\boldsymbol{\sigma}(\mathbf{x}) \mathbf{n}(\mathbf{x}) &= \mathbf{0}, \quad &&\mathbf{x} \in \partial\Omega_\mathrm{wavy},
\end{alignat} \label{eq:HalfStarMatrixBCs}%
\end{subequations}
and the displacement $\mathbf{u}(\mathbf{x})$ is measured at locations $\partial \Omega^m$ distributed along the exposed wavy boundary $\partial\Omega_\mathrm{wavy}$.

For the 3D square elastic matrix (Fig.~6, middle row), the domain is $\Omega = [-0.5,0.5] \times [-0.5,0.5] \times [-0.5,0.5]$. The boundary conditions are
\begin{subequations}
\begin{alignat}{2}
\boldsymbol{\sigma}(\mathbf{x}) \mathbf{n}(\mathbf{x}) &= \pm P_o \mathbf{e}_1, &\quad &\mathbf{x} \in \{\pm 0.5\} \times [-0.5,0.5] \times [-0.5,0.5], \\
\boldsymbol{\sigma}(\mathbf{x}) \mathbf{n}(\mathbf{x}) &= \mathbf{0}, \quad &&\mathbf{x} \in [-0.5,0.5] \times \{\pm 0.5\} \times [-0.5,0.5], \\
\boldsymbol{\sigma}(\mathbf{x}) \mathbf{n}(\mathbf{x}) &= \mathbf{0}, \quad &&\mathbf{x} \in [-0.5,0.5] \times [-0.5,0.5] \times \{\pm 0.5\},
\end{alignat} \label{eq:3DMatrixBCs}%
\end{subequations}
and the displacement $\mathbf{u}(\mathbf{x})$ is measured at locations $\partial \Omega^m$ distributed along the entire external boundary $\partial \Omega$.

For the elastic layer (Fig.~1b), the domain is $\Omega = [0,1] \times [-0.5,0]$. The boundary conditions are
\begin{subequations}
\begin{alignat}{2}
\boldsymbol{\sigma}(\mathbf{x}) \mathbf{n}(\mathbf{x}) &= -P_o \mathbf{e}_2, &\quad& \mathbf{x} \in [0,1] \times \{0\}, \\
\mathbf{u} &= \mathbf{0}, && \mathbf{x} \in [0,1] \times \{-0.5\},
\end{alignat} \label{eq:LayerBCs}%
\end{subequations}
in addition to periodic displacement and traction boundary conditions on $\mathbf{x} \in \{0,1\} \times [-0.5,0]$. The displacement $\mathbf{u}(\mathbf{x})$ is measured at locations $\partial \Omega^m$ distributed along the top surface $\partial \Omega_t = [0,1] \times \{0\}$.


\subsubsection{Large-deformation nonlinear hyperelasticity} 
\label{app:LargeDeformationNonlinearHyperelasticity}

\textbf{Physical quantities.} Next, we consider the case where the elastic body consists of an incompressible Neo-Hookean hyperelastic material with shear modulus $\mu$.
We now have to distinguish between the reference (undeformed) and current (deformed) configurations. 
We denote by $\mathbf{x} = (x_1,x_2) \in \Omega$ and $\mathbf{y} = (y_1,y_2) \in \Omega^*$ the coordinates in the reference and deformed configurations, respectively, with $\Omega^*$ the deformed image of $\Omega$. 
The displacement field $\mathbf{u}(\mathbf{x})$ with components $u_i(\mathbf{x})$ moves an initial position $\mathbf{x} \in \Omega$ into its current location $\mathbf{y} = \mathbf{x} + \mathbf{u}(\mathbf{x}) \in \Omega^*$. 
In order to formulate the governing equations and boundary conditions in the reference configuration $\Omega$, we need to introduce the first Piola-Kirchhoff stress tensor $\mathbf{S}(\mathbf{x})$ with components $S_{ij}(\mathbf{x})$. Unlike the Cauchy stress tensor, the first Piola-Kirchhoff stress tensor is defined in $\Omega$ and is not symmetric. The deformation of the elastic body is then described by the vector field $\boldsymbol{\psi}(\mathbf{x}) = (\mathbf{u}(\mathbf{x}), \mathbf{S}(\mathbf{x}),p(\mathbf{x}))$ defined over $\Omega$, where $p(\mathbf{x})$ is a pressure field that serves to enforce the incompressibility constraint.

\textbf{Governing PDEs.} The equilibrium equations are
\begin{equation}
\sum_{j}\frac{\partial S_{ij}}{\partial x_j} = 0, \quad \mathbf{x} \in \Omega,
\label{eq:NonlinearEquilibrium}
\end{equation}
where the derivatives in $\nabla_\mathbf{x}$ are taken with respect to the reference coordinates $\mathbf{x}$. We only consider the presence of voids so that the nonlinear constitutive law $F(\mathbf{S}, \nabla_\mathbf{x} \mathbf{u}, p, \rho) = 0$ is simply expressed as
\begin{equation}
\mathbf{S} = \rho \left[ -p \mathbf{F}^{-T} + \mu \mathbf{F} \right], \quad \mathbf{x} \in \Omega,
\label{eq:NonlinearStressStrain}
\end{equation}
where $\mathbf{F}(\mathbf{x}) = \mathbf{I} + \nabla_\mathbf{x} \mathbf{u}(\mathbf{x})$ is the deformation gradient tensor. Notice that the stress vanishes in the $\rho = 0$ regions. Finally, we have the incompressibility constraint
\begin{equation}
\rho \left[ \det(\mathbf{F}) - 1 \right] = 0, \quad \mathbf{x} \in \Omega,
\label{eq:NonlinearIncompressibility}
\end{equation}
which turns itself off in the $\rho = 0$ regions since voids do not deform in a way that preserves volume.

\textbf{Boundary conditions.} We only treat the matrix problem (Fig.~1a) in this hyperelastic case. The domain is $\Omega = [-0.5,0.5] \times [-0.5,0.5]$ and the boundary conditions are
\begin{subequations}
\begin{alignat}{2}
\mathbf{S}(\mathbf{x}) \mathbf{n}_0(\mathbf{x}) &= -P_o \mathbf{e}_1, &\quad &\mathbf{x} \in \{-0.5,0.5\} \times [-0.5,0.5], \\
\mathbf{S}(\mathbf{x}) \mathbf{n}_0(\mathbf{x}) &= \mathbf{0}, \quad &&\mathbf{x} \in [-0.5,0.5] \times \{-0.5,0.5\},
\end{alignat} \label{eq:MatrixBCsHyperEla}%
\end{subequations}
where $P_o/E = 0.173$. As in the linear elastic case, the displacement $\mathbf{u}(\mathbf{x})$ is measured at locations $\partial \Omega^m$ distributed along the entire external boundary $\partial \Omega$.


\subsection{Thermal conduction experiments} 
\label{app:ThermalExperiments}

\textbf{Physical quantities.} In heat transfer problems, the response of the body is described by the vector field $\boldsymbol{\psi}(\mathbf{x}) = (T(\mathbf{x}), \mathbf{q}(\mathbf{x}))$, where $T(\mathbf{x})$ is the temperature field and $\mathbf{q}(\mathbf{x})$ is the heat flux with components $q_i(\mathbf{x})$. Since we only consider 2D thermal imaging cases, the index $i$ will hereafter range from 1 to 2.

\textbf{Governing PDEs.} The governing PDEs comprise the heat conservation law stipulating that, at steady-state,
\begin{equation}
\sum_{i}\frac{\partial q_i}{\partial x_i} = 0, \quad \mathbf{x} \in \Omega,
\label{eq:HeatConservation}
\end{equation}
as well as Fourier's law of heat conduction relating the heat flux and the temperature. For cases involving perfectly insulating inclusions, we express Fourier's law as
\begin{equation}
\mathbf{q} + \rho k(T) \nabla T = 0, \quad \mathbf{x} \in \Omega,
\label{eq:FourierLaw}
\end{equation}
while for cases involving perfectly conductive inclusions, we express Fourier's law as
\begin{equation}
\rho \mathbf{q} + k(T) \nabla T = 0, \quad \mathbf{x} \in \Omega.
\label{eq:InvertedFourierLaw}
\end{equation}
In both expressions, $k$ denotes the thermal conductivity of the material, which may be a function of temperature. For case 26 in Tab.~III, we assume that the matrix is made of steel and therefore consider a constant thermal conductivity $k = 45 \, \mathrm{W}/\mathrm{mK}$, resulting in a linear PDE. For cases 27 and 28 in Tab.~III, we assume that the matrix is made of a material with temperature-dependent thermal conductivity $k(T) = k_0(1+T/T_0)$, where $k_0 = 1 \, \mathrm{W}/\mathrm{mK}$ is a reference thermal conductivity and $T_0 = 1 \, \mathrm{K}$ is a reference temperature, resulting in a nonlinear PDE. Notice that \eqref{eq:FourierLaw} and \eqref{eq:InvertedFourierLaw} are formulated such that when $\rho = 0$, the heat flux vanishes for perfectly insulating inclusions, while the temperature becomes uniform for perfectly conductive inclusions.

\textbf{Boundary conditions.} For all thermal cases considered in this paper, the domain is $\Omega = [-0.5,0.5] \times [-0.5,0.5]$. For case 26 in Tab.~III, the left side of the body is heated to a prescribed temperature $T_0 = 478 \, \mathrm{K}$, while the three remaining sides are left exposed to air at ambient temperature $T_a = 278 \, \mathrm{K}$. The corresponding boundary conditions are
\begin{subequations}
\begin{alignat}{2}
T(\mathbf{x}) &= T_0, &\quad &\mathbf{x} \in \{ - 0.5 \} \times [-0.5,0.5], \\
\mathbf{q}(\mathbf{x}) \cdot \mathbf{n}(\mathbf{x}) &= \epsilon \sigma (T^4 - T_a^4), \quad &&\mathbf{x} \in \{ 0.5 \} \times [-0.5,0.5] \cup [-0.5,0.5] \times \{ \pm 0.5 \}.
\end{alignat} \label{eq:MatrixBCsThermal1}%
\end{subequations}
where the three exposed sides have been modeled using a radiation boundary condition that involves the emissivity $\epsilon = 0.9$ of a rough steel surface as well as the Stefan-Boltzmann constant $\sigma = 5.67 \cdot 10^{-8} \, \mathrm{W}/(\mathrm{m}^2 \mathrm{K}^4)$. The temperature $T(\mathbf{x})$ is measured at locations $\partial \Omega^m$ distributed along the three exposed sides.

For cases 27 and 28 in Tab.~III, the boundary conditions are
\begin{subequations}
\begin{alignat}{2}
T(\mathbf{x}) &= T_0, &\quad &\mathbf{x} \in \{ - 0.5 \} \times [-0.5,0.5], \\
T(\mathbf{x}) &= T_1, &\quad &\mathbf{x} \in \{ 0.5 \} \times [-0.5,0.5], \\
\mathbf{q}(\mathbf{x}) \cdot \mathbf{n}(\mathbf{x}) &= 0, \quad &&\mathbf{x} \in [-0.5,0.5] \times \{ \pm 0.5 \}.
\end{alignat} \label{eq:MatrixBCsThermal2}%
\end{subequations}
where $T_0 = 1 \, \mathrm{K}$ and $T_1 = 0 \, \mathrm{K}$. These cases are repeated four times, in each instance assuming that both the applied boundary condition and measurements are unavailable on one of the four sides, while the temperature $T(\mathbf{x})$ or normal heat flux $\mathbf{q}(\mathbf{x}) \cdot \mathbf{n}(\mathbf{x})$ resulting from the prescribed thermal loading are measured at locations $\partial \Omega^m$ distributed along the remaining three sides.

\section{Detailed solution methodology}
\label{app:AdditionalInformationSolutionMethodology}

\subsection{Mechanical experiments}

\subsubsection{Small-deformation linear elasticity}

We describe the implementation of the 2D cases, which is easily generalized to 3D. Since the physical quantities are $\boldsymbol{\psi} = (u_1,u_2,\sigma_{11},\sigma_{22},\sigma_{12})$, we introduce the neural network approximations
\begin{subequations}
\begin{align}
u_1(\mathbf{x}) &= \bar{u}_1(\mathbf{x}; \boldsymbol{\theta}_1), \\
u_2(\mathbf{x}) &= \bar{u}_2(\mathbf{x}; \boldsymbol{\theta}_2), \\
\sigma_{11}(\mathbf{x}) &= \bar{\sigma}_{11}(\mathbf{x}; \boldsymbol{\theta}_3), \\
\sigma_{22}(\mathbf{x}) &= \bar{\sigma}_{22}(\mathbf{x}; \boldsymbol{\theta}_4), \\
\sigma_{12}(\mathbf{x}) &= \bar{\sigma}_{12}(\mathbf{x}; \boldsymbol{\theta}_5), \\
\phi(\mathbf{x}) &= \bar{\phi}(\mathbf{x}; \boldsymbol{\theta}_\phi).
\end{align} \label{eq:LinearNN}%
\end{subequations}
The last equation represents the level-set neural network, which defines the material density as $\rho(\mathbf{x}) = \bar{\rho}(\mathbf{x};\boldsymbol{\theta}_\phi) = \mathrm{sigmoid}(\bar{\phi}(\mathbf{x};\boldsymbol{\theta}_\phi)/\delta)$. We then formulate the loss function $\mathcal{L}$ by specializing the loss term expressions described in the main text to the linear elasticity problem, using the governing equations given in Appendix \ref{app:SmallDeformationLinearElasticity}. Omitting the $\boldsymbol{\theta}$'s for notational simplicity, we obtain
\begin{subequations}
\begin{align}
\mathcal{L}_\mathrm{meas}(\boldsymbol{\theta}_{\boldsymbol{\psi}}) &= \frac{1}{|\partial \Omega^m|} \sum_{\mathbf{x}_i \in \partial \Omega^m}  |\bar{\mathbf{u}}(\mathbf{x}_i) - \mathbf{u}_i^m|^2, \label{eq:LinearLmeas} \\
\mathcal{L}_\mathrm{gov}(\boldsymbol{\theta}_{\boldsymbol{\psi}}, \boldsymbol{\theta}_\phi) &= \frac{1}{|\Omega^d|} \sum_{\mathbf{x}_i \in \Omega^d} |\mathbf{r}_\mathrm{eq}(\bar{\boldsymbol{\sigma}}(\mathbf{x}_i))|^2 + \frac{1}{|\Omega^d|} \sum_{\mathbf{x}_i \in \Omega^d} |\mathbf{r}_\mathrm{cr}(\bar{\mathbf{u}}(\mathbf{x}_i),\bar{\boldsymbol{\sigma}}(\mathbf{x}_i),\bar{\rho}(\mathbf{x}_i))|^2, \label{eq:LinearLF},
\end{align}
\end{subequations}
where $\bar{\mathbf{u}} = (\bar{u}_1,\bar{u}_2)$ and $\bar{\boldsymbol{\sigma}}$ has components $\bar{\sigma}_{i,j}$, $i,j=1,2$. In \eqref{eq:LinearLF}, the terms $\mathbf{r}_\mathrm{eq}$ and $\mathbf{r}_\mathrm{cr}$ refer to the residuals of the equilibrium equation \eqref{eq:LinearEquilibrium} and the constitutive relation \eqref{eq:LinearStressStrainVoid}, \eqref{eq:LinearStressStrainInclusion2D}, or \eqref{eq:LinearStressStrainInclusion3D}. The eikonal loss term is problem-independent and therefore identical to the expression given in (6) in the main text.

We note that instead of defining neural network approximations for the displacements and the stresses, we could define neural network approximations solely for the displacements, that is, $\boldsymbol{\psi} = (u_1,u_2)$. In this case, the loss term \eqref{eq:LinearLF} would only include the residual of the equilibrium equation \eqref{eq:LinearEquilibrium}, in which the stress components would be directly expressed in terms of the displacements and the material distribution using the constitutive relation \eqref{eq:LinearStressStrainVoid}. However, several recent studies \citep{rao2021,haghighat2021,henkes2022,rezaei2022,gladstone2022,harandi2023} have shown that the mixed formulation adopted in the present work results in superior accuracy and training performance, which could partly be explained by the fact that only first-order derivatives of the neural network outputs are involved since the displacements and stresses are only differentiated to first oder in \eqref{eq:LinearEquilibrium} and \eqref{eq:LinearStressStrainVoid}. In our case, the mixed formulation holds the additional advantage that it enables us to treat stiff and rigid inclusions using the inverted constitutive relation \eqref{eq:LinearStressStrainInclusion2D} instead of \eqref{eq:LinearStressStrainVoid}. Finally, the mixed formulation allows us to directly integrate both displacement and traction boundary conditions into the output of the neural network approximations, as we describe in the next paragraph.

We design the architecture of the neural networks in such a way that they inherently satisfy the boundary conditions, treating the latter as hard constraints \citep{dong2021,sukumar2022}. For the square elastic matrix, we do this through the transformations
\begin{subequations}
\begin{align}
\bar{u}_1(\mathbf{x}; \boldsymbol{\theta}_1) &= \bar{u}_1'(\mathbf{x}; \boldsymbol{\theta}_1), \\
\bar{u}_2(\mathbf{x}; \boldsymbol{\theta}_2) &= \bar{u}_2'(\mathbf{x}; \boldsymbol{\theta}_2), \\
\bar{\sigma}_{11}(\mathbf{x}; \boldsymbol{\theta}_3) &= (x-0.5)(x+0.5) \, \bar{\sigma}_{11}'(\mathbf{x}; \boldsymbol{\theta}_3) + P_o, \\
\bar{\sigma}_{22}(\mathbf{x}; \boldsymbol{\theta}_4) &= (y-0.5)(y+0.5) \, \bar{\sigma}_{22}'(\mathbf{x}; \boldsymbol{\theta}_4), \\
\bar{\sigma}_{12}(\mathbf{x}; \boldsymbol{\theta}_5) &= (x-0.5)(x+0.5)(y-0.5)(y+0.5) \, \bar{\sigma}_{12}'(\mathbf{x}; \boldsymbol{\theta}_5), \\
\bar{\phi}(\mathbf{x}; \boldsymbol{\theta}_\phi) &= (x-0.5)(x+0.5)(y-0.5)(y+0.5) \, \bar{\phi}'(\mathbf{x}; \boldsymbol{\theta}_\phi) + w,
\end{align}
\end{subequations}
where the quantities with a prime denote the raw output of the neural network. In this way, the neural network approximations defined in \eqref{eq:LinearNN} obey by construction the boundary conditions \eqref{eq:2DMatrixBCs}. Further, since we know that the elastic material is present all along the outer surface $\partial \Omega$, we define $\bar{\phi}$ so that $\rho = \mathrm{sigmoid}(\bar{\phi}/\delta) \simeq 1$ on $\partial \Omega$ (recall that $w = 10 \delta$). In the case of the M, I, T inclusions, these transformations are trivially changed to reflect the fact that the matrix is wider and pulled from the top and bottom. 

The half-star elastic matrix calls for more complex transformations due to its wavy boundary. Let
\begin{subequations}
\begin{align}
\bar{u}_1(\mathbf{x}; \boldsymbol{\theta}_1) &= y \, \bar{u}_1'(\mathbf{x}; \boldsymbol{\theta}_1), \\
\bar{u}_2(\mathbf{x}; \boldsymbol{\theta}_2) &= y \, \bar{u}_2'(\mathbf{x}; \boldsymbol{\theta}_2), \\
\bar{\sigma}_{11}(\mathbf{x}; \boldsymbol{\theta}_3, \boldsymbol{\theta}_6) &= \Phi(\mathbf{x}) \bar{\sigma}_{11}'(\mathbf{x}; \boldsymbol{\theta}_3) + f_{11}(\mathbf{x}) \bar{\sigma}_{mm}'(\mathbf{x}; \boldsymbol{\theta}_6), \label{eq:HalfStarStress11} \\
\bar{\sigma}_{22}(\mathbf{x}; \boldsymbol{\theta}_4, \boldsymbol{\theta}_6) &= \Phi(\mathbf{x}) \bar{\sigma}_{22}'(\mathbf{x}; \boldsymbol{\theta}_4) + f_{22}(\mathbf{x}) \bar{\sigma}_{mm}'(\mathbf{x}; \boldsymbol{\theta}_6), \\
\bar{\sigma}_{12}(\mathbf{x}; \boldsymbol{\theta}_5, \boldsymbol{\theta}_6) &= \Phi(\mathbf{x}) \bar{\sigma}_{12}'(\mathbf{x}; \boldsymbol{\theta}_5) + f_{12}(\mathbf{x}) \bar{\sigma}_{mm}'(\mathbf{x}; \boldsymbol{\theta}_6), \label{eq:HalfStarStress12} \\
\bar{\phi}(\mathbf{x}; \boldsymbol{\theta}_\phi) &= y \Phi(\mathbf{x}) \, \bar{\phi}'(\mathbf{x}; \boldsymbol{\theta}_\phi) + w,
\end{align}
\end{subequations}
where $\Phi(\mathbf{x})$ is a function that smoothly vanishes on the wavy boundary $\partial \Omega_\mathrm{wavy}$ and is positive elsewhere. Furthermore, $f_{ij}(\mathbf{x}) = \Psi(\mathbf{x}) m_i(\mathbf{x}) m_j(\mathbf{x})$ for $i,j = 1,2$, where $\Psi(\mathbf{x})$ is a function that equates 1 on $\partial \Omega_\mathrm{wavy}$ and smoothly vanishes elsewhere, and $m_i(\mathbf{x})$ are smooth functions equal on $\partial \Omega_\mathrm{wavy}$ to the components of the tangent unit vector. That way, the neural network approximations defined in \eqref{eq:LinearNN} obey by construction the boundary conditions \eqref{eq:HalfStarMatrixBCs}. Concretely, we use the expressions
\begin{subequations}
\begin{align}
\Phi(\mathbf{x}) &= 1 - \left( \frac{r}{r^* + a^* \sin 7 \theta} \right)^2, \\
\Psi(\mathbf{x}) &= \frac{1 + \cos ( \pi \min(|\Phi(\mathbf{x})/\Phi_c|, 1) )}{2}, \\
m_1(\mathbf{x}) &= 7 a^* \cos 7 \theta \cos \theta - (r^* + a^* \sin 7 \theta) \sin \theta, \\
m_2(\mathbf{x}) &= 7 a^* \cos 7 \theta \sin \theta + (r^* + a^* \sin 7 \theta) \cos \theta.
\end{align}
\end{subequations}
Note that the transformations defined in \eqref{eq:HalfStarStress11}-\eqref{eq:HalfStarStress12} require the introduction of an additional neural network $\bar{\sigma}_{mm}'$ with parameters $\boldsymbol{\theta}_6$ that represents the tangential stress along $\partial \Omega_\mathrm{wavy}$. Finally, since we know that the elastic material is present all along the outer perimeter $\partial \Omega$, we define $\bar{\phi}$ so that $\rho = \mathrm{sigmoid}(\bar{\phi}/\delta) \simeq 1$ on $\partial \Omega$.

For the periodic elastic layer, we introduce the transformations
\begin{subequations}
\begin{align}
\bar{u}_1(\mathbf{x}; \boldsymbol{\theta}_1) &= (y+0.5) \, \bar{u}_1'(\cos x, \sin x, y; \boldsymbol{\theta}_1), \\
\bar{u}_2(\mathbf{x}; \boldsymbol{\theta}_2) &= (y+0.5) \, \bar{u}_2'(\cos x, \sin x, y; \boldsymbol{\theta}_2), \\
\bar{\sigma}_{11}(\mathbf{x}; \boldsymbol{\theta}_3) &= \bar{\sigma}_{11}'(\cos x, \sin x, y; \boldsymbol{\theta}_3), \\
\bar{\sigma}_{22}(\mathbf{x}; \boldsymbol{\theta}_4) &= y \, \bar{\sigma}_{22}'(\cos x, \sin x, y; \boldsymbol{\theta}_4) - P_o, \\
\bar{\sigma}_{12}(\mathbf{x}; \boldsymbol{\theta}_5) &= y \, \bar{\sigma}_{12}'(\cos x, \sin x, y; \boldsymbol{\theta}_5), \\
\bar{\phi}(\mathbf{x}; \boldsymbol{\theta}_\phi) &= y(y+0.5) \, \bar{\phi}'(\cos x, \sin x, y; \boldsymbol{\theta}_\phi) + w(4y+1),
\end{align}
\end{subequations}
so that the neural network approximations defined in \eqref{eq:LinearNN} obey by construction the boundary conditions \eqref{eq:LayerBCs} and are periodic along the $x$ direction. Further, since we know that the elastic material is present all along the top surface $y = 0$ and the rigid substrate is present all along the bottom surface $y = -0.5$, we define $\bar{\phi}$ so that $\rho = \mathrm{sigmoid}(\bar{\phi}/\delta) \simeq 1$ for $y = 0$ and $\rho \simeq 0$ for $y=-0.5$.

\subsubsection{Large-deformation hyperelasticity}

The problem is now described by the physical quantities $\boldsymbol{\psi} = (u_1,u_2,S_{11},S_{22},S_{12},S_{21},p)$. We therefore introduce the neural network approximations
\begin{subequations}
\begin{align}
u_1(\mathbf{x}) &= \bar{u}_1(\mathbf{x}; \boldsymbol{\theta}_1), \\
u_2(\mathbf{x}) &= \bar{u}_2(\mathbf{x}; \boldsymbol{\theta}_2), \\
S_{11}(\mathbf{x}) &= \bar{S}_{11}(\mathbf{x}; \boldsymbol{\theta}_3), \\
S_{22}(\mathbf{x}) &= \bar{S}_{22}(\mathbf{x}; \boldsymbol{\theta}_4), \\
S_{12}(\mathbf{x}) &= \bar{S}_{12}(\mathbf{x}; \boldsymbol{\theta}_5), \\
S_{21}(\mathbf{x}) &= \bar{S}_{21}(\mathbf{x}; \boldsymbol{\theta}_6), \\
p(\mathbf{x}) &= \bar{p}(\mathbf{x}; \boldsymbol{\theta}_7), \\
\phi(\mathbf{x}) &= \bar{\phi}(\mathbf{x}; \boldsymbol{\theta}_\phi),
\end{align} \label{eq:NonlinearNN}%
\end{subequations}
and the material distribution is given by $\rho(\mathbf{x}) = \bar{\rho}(\mathbf{x};\boldsymbol{\theta}_\phi) = \mathrm{sigmoid}(\bar{\phi}(\mathbf{x};\boldsymbol{\theta}_\phi)/\delta)$. We then formulate the loss function $\mathcal{L}$ by specializing the loss term expressions described in the main text to the hyperelasticity problem, using the governing equations given in Appendix \ref{app:LargeDeformationNonlinearHyperelasticity}. Omitting the $\boldsymbol{\theta}$'s for notational simplicity, we obtain
\begin{subequations}
\begin{align}
\mathcal{L}_\mathrm{meas}(\boldsymbol{\theta}_{\boldsymbol{\psi}}) &= \frac{1}{|\partial \Omega^m|} \sum_{\mathbf{x}_i \in \partial \Omega^m}  |\bar{\mathbf{u}}(\mathbf{x}_i) - \mathbf{u}_i^m|^2, \label{eq:NonlinearLmeas} \\
\mathcal{L}_\mathrm{gov}(\boldsymbol{\theta}_{\boldsymbol{\psi}}, \boldsymbol{\theta}_\phi) &= \frac{1}{|\Omega^d|} \sum_{\mathbf{x}_i \in \Omega^d} |\mathbf{r}_\mathrm{eq}(\bar{\mathbf{S}}(\mathbf{x}_i))|^2 + \frac{1}{|\Omega^d|} \sum_{\mathbf{x}_i \in \Omega^d} |\mathbf{r}_\mathrm{cr}(\bar{\mathbf{u}}(\mathbf{x}_i),\bar{\mathbf{S}}(\mathbf{x}_i),\bar{p}(\mathbf{x}_i),\bar{\rho}(\mathbf{x}_i))|^2 \nonumber \\
&\quad+ \frac{1}{|\Omega^d|} \sum_{\mathbf{x}_i \in \Omega^d} |\mathbf{r}_\mathrm{inc}(\bar{\mathbf{u}}(\mathbf{x}_i),\bar{\rho}(\mathbf{x}_i))|^2, \label{eq:NonlinearLF}
\end{align}
\end{subequations}
where $\bar{\mathbf{u}} = (\bar{u}_1,\bar{u}_2)$ and $\bar{\mathbf{S}}$ has components $\bar{S}_{i,j}$, $i,j=1,2$. In \eqref{eq:NonlinearLF}, the terms $\mathbf{r}_\mathrm{eq}$, $\mathbf{r}_\mathrm{cr}$, and $\mathbf{r}_\mathrm{inc}$ refer to the residuals of the equilibrium equation \eqref{eq:NonlinearEquilibrium}, the constitutive relation \eqref{eq:NonlinearStressStrain}, and the incompressibility constraint \eqref{eq:NonlinearIncompressibility}. The eikonal loss term is problem-independent and therefore identical to the expression given in (6) in the main text.

As in the linear elasticity case, we design the architecture of the neural networks in such a way that they inherently satisfy the boundary conditions. For the elastic matrix problem,
\begin{subequations}
\begin{align}
\bar{u}_1(\mathbf{x}; \boldsymbol{\theta}_1) &= \bar{u}_1'(\mathbf{x}; \boldsymbol{\theta}_1), \\
\bar{u}_2(\mathbf{x}; \boldsymbol{\theta}_2) &= \bar{u}_2'(\mathbf{x}; \boldsymbol{\theta}_2), \\
\bar{S}_{11}(\mathbf{x}; \boldsymbol{\theta}_3) &= (x-0.5)(x+0.5) \, \bar{S}_{11}'(\mathbf{x}; \boldsymbol{\theta}_3) + P_o, \\
\bar{S}_{22}(\mathbf{x}; \boldsymbol{\theta}_4) &= (y-0.5)(y+0.5) \, \bar{S}_{22}'(\mathbf{x}; \boldsymbol{\theta}_4), \\
\bar{S}_{12}(\mathbf{x}; \boldsymbol{\theta}_5) &= (y-0.5)(y+0.5) \, \bar{S}_{12}'(\mathbf{x}; \boldsymbol{\theta}_5), \\
\bar{S}_{21}(\mathbf{x}; \boldsymbol{\theta}_6) &= (x-0.5)(x+0.5) \, \bar{S}_{21}'(\mathbf{x}; \boldsymbol{\theta}_6), \\
\bar{p}(\mathbf{x}; \boldsymbol{\theta}_7) &= \bar{p}'(\mathbf{x}; \boldsymbol{\theta}_7), \\
\bar{\phi}(\mathbf{x}; \boldsymbol{\theta}_\phi) &= (x-0.5)(x+0.5)(y-0.5)(y+0.5) \, \bar{\phi}'(\mathbf{x}; \boldsymbol{\theta}_\phi) + w,
\end{align}
\end{subequations}
where the quantities with a prime denote the raw output of the neural network. In this way, the neural network approximations defined in \eqref{eq:NonlinearNN} obey by construction the boundary conditions \eqref{eq:MatrixBCsHyperEla}. As before, since we know that the elastic material is present all along the outer surface $\partial \Omega$, we define $\bar{\phi}$ so that $\phi = w$ on $\partial \Omega$, which ensures that $\rho = \mathrm{sigmoid}(\phi/\delta) \simeq 1$ on $\partial \Omega$.

\subsection{Thermal conduction experiments}

Since the physical quantities are $\boldsymbol{\psi} = (T, q_1, q_2)$, we introduce the neural network approximations
\begin{subequations}
\begin{align}
T(\mathbf{x}) &= T(\mathbf{x}; \boldsymbol{\theta}_1), \\
q_1(\mathbf{x}) &= \bar{q}_1(\mathbf{x}; \boldsymbol{\theta}_2), \\
q_2(\mathbf{x}) &= \bar{q}_2(\mathbf{x}; \boldsymbol{\theta}_3), \\
\phi(\mathbf{x}) &= \bar{\phi}(\mathbf{x}; \boldsymbol{\theta}_\phi).
\end{align}%
\end{subequations}
The last equation represents the level-set neural network, which defines the material density as $\rho(\mathbf{x}) = \bar{\rho}(\mathbf{x};\boldsymbol{\theta}_\phi) = \mathrm{sigmoid}(\bar{\phi}(\mathbf{x};\boldsymbol{\theta}_\phi)/\delta)$. We then formulate the loss function $\mathcal{L}$ by specializing the loss term expressions described in the main text to the heat transfer problem, using the governing equations given in Appendix \ref{app:ThermalExperiments}. Omitting the $\boldsymbol{\theta}$'s for notational simplicity, we obtain
\begin{subequations}
\begin{align}
\mathcal{L}_\mathrm{meas}(\boldsymbol{\theta}_{\boldsymbol{\psi}}) &= \frac{1}{|\partial \Omega_T^m|} \sum_{\mathbf{x}_i \in \partial \Omega_T^m}  |\bar{T}(\mathbf{x}_i) - T_i^m|^2 + \frac{1}{|\partial \Omega_q^m|} \sum_{\mathbf{x}_i \in \partial \Omega_q^m}  |\bar{\mathbf{q}}(\mathbf{x}_i) \cdot \mathbf{n} - q_i^m|^2, \label{eq:ThermalLmeas} \\
\mathcal{L}_\mathrm{gov}(\boldsymbol{\theta}_{\boldsymbol{\psi}}, \boldsymbol{\theta}_\phi) &= \frac{1}{|\Omega^d|} \sum_{\mathbf{x}_i \in \Omega^d} |\mathbf{r}_\mathrm{hc}(\bar{T}(\mathbf{x}_i))|^2 + \frac{1}{|\Omega^d|} \sum_{\mathbf{x}_i \in \Omega^d} |\mathbf{r}_\mathrm{fl}(\bar{T}(\mathbf{x}_i),\bar{\mathbf{q}}(\mathbf{x}_i),\bar{\rho}(\mathbf{x}_i))|^2, \label{eq:ThermalLF},
\end{align}
\end{subequations}
where $\bar{\mathbf{q}} = (\bar{q}_1,\bar{q}_2)$. In \eqref{eq:ThermalLmeas}, $\partial \Omega_T^m$ and $\partial \Omega_q^m$ denote portions of the boundary with temperature and normal heat flux measurements, respectively. In \eqref{eq:ThermalLF}, the terms $\mathbf{r}_\mathrm{hc}$ and $\mathbf{r}_\mathrm{fl}$ refer to the residuals of the heat conservation equation \eqref{eq:HeatConservation} and Fourier's law \eqref{eq:FourierLaw} or \eqref{eq:InvertedFourierLaw}. The eikonal loss term is problem-independent and therefore identical to the expression given in (6) in the main text.

As with the other examples, we design the architecture of the neural networks in such a way that they inherently satisfy the boundary conditions. For case 26 in Tab.~III, we do this through the transformations
\begin{subequations}
\begin{align}
\bar{T}(\mathbf{x}; \boldsymbol{\theta}_1) &= (x+0.5) \, \bar{T}'(\mathbf{x}; \boldsymbol{\theta}_1) + T_0, \\
\bar{q}_1(\mathbf{x}; \boldsymbol{\theta}_2) &= (x-0.5) \, \bar{q}_1'(\mathbf{x}; \boldsymbol{\theta}_2) + q_n(\mathbf{x}), \\
\bar{q}_2(\mathbf{x}; \boldsymbol{\theta}_3) &= (y-0.5)(y+0.5) \, \bar{q}_2'(\mathbf{x}; \boldsymbol{\theta}_3) + (y-0.5) \, q_n(\mathbf{x}) + (y+0.5) \, q_n(\mathbf{x}), \\
\bar{\phi}(\mathbf{x}; \boldsymbol{\theta}_\phi) &= (x-0.5)(x+0.5)(y-0.5)(y+0.5) \, \bar{\phi}'(\mathbf{x}; \boldsymbol{\theta}_\phi) + w,
\end{align}
\end{subequations}
where $q_n(\mathbf{x}) = \epsilon \sigma (\mathtt{stop\_gradient}\{\bar{T}(\mathbf{x}; \boldsymbol{\theta}_1)\}^4 - T_a^4)$, and the quantities with a prime denote the raw output of the neural network. In this way, the neural network approximations defined in \eqref{eq:LinearNN} obey by construction the boundary conditions \eqref{eq:MatrixBCsThermal1}. Further, since we know that the elastic material is present all along the outer surface $\partial \Omega$, we define $\bar{\phi}$ so that $\rho = \mathrm{sigmoid}(\bar{\phi}/\delta) \simeq 1$ on $\partial \Omega$ (recall that $w = 10 \delta$).

For cases 27 and 28 in Tab.~III, we utilize the transformations
\begin{subequations}
\begin{align}
\bar{T}(\mathbf{x}; \boldsymbol{\theta}_1) &= (x-0.5)(x+0.5) \, \bar{T}'(\mathbf{x}; \boldsymbol{\theta}_1) - (x-0.5) \, T_0, \\
\bar{q}_1(\mathbf{x}; \boldsymbol{\theta}_2) &= \bar{q}_1'(\mathbf{x}; \boldsymbol{\theta}_2), \\
\bar{q}_2(\mathbf{x}; \boldsymbol{\theta}_3) &= (y-0.5)(y+0.5) \, \bar{q}_2'(\mathbf{x}; \boldsymbol{\theta}_3), \\
\bar{\phi}(\mathbf{x}; \boldsymbol{\theta}_\phi) &= (x-0.5)(x+0.5)(y-0.5)(y+0.5) \, \bar{\phi}'(\mathbf{x}; \boldsymbol{\theta}_\phi) + w,
\end{align}
\end{subequations}
where the quantities with a prime denote the raw output of the neural network. In this way, the neural network approximations defined in \eqref{eq:LinearNN} obey by construction the boundary conditions \eqref{eq:MatrixBCsThermal2}. Note, however, that these transformations are modified to reflect the fact that the applied boundary condition is presumed unknown on one of the four sides. Further, since we know that the elastic material is present all along the outer surface $\partial \Omega$, we define $\bar{\phi}$ so that $\rho = \mathrm{sigmoid}(\bar{\phi}/\delta) \simeq 1$ on $\partial \Omega$ (recall that $w = 10 \delta$).

\subsection{Rescaling} 
\label{app:Rescaling}

The physical quantities involved in these mechanical and heat transfer problems span a wide range of scales; for instance, displacements may be orders of magnitude smaller than the length scale associated with the geometry. In order to obtain balanced gradients between the different components of the loss function during training, we rescale all physical quantities into nondimensional values of order one and implement the corresponding nondimensional equations, as also done, for example, in \cite{henkes2022}. In mechanical experiments (all cases in Tabs.~I and II), lengths are rescaled with the characteristic length $L$ of the geometry, tractions and stresses with the magnitude $P_o$ of the applied traction at the boundaries, and displacements with the ratio $L P_o / E$, where $E$ is the Young's modulus of the elastic material (in the hyperelastic case, we use the equivalent Young's modulus $E = 3 \mu$, where $\mu$ is the shear modulus of the hyperelastic material). In the thermal conduction experiment corresponding to case 26 in Tab.~III, we express the temperature through the deviation $\Delta T = T - T_a$, which we rescale with $T_0 - T_a$, and we rescale the heat flux with $\varepsilon \sigma (T_0^4 - T_a^4)$. In the thermal conduction experiment corresponding to cases 27 and 28 in Tab.~III, the temperature is rescaled with $T_0$ and the heat flux with $k T_0/L$.

\section{Architecture and training details}
\label{app:ImplementationDetails}

This section provides details on the architecture of the deep neural networks, the training procedure, and the parameter values considered in this study.

\subsection{Neural network architecture}
\label{app:NeuralNetworkFormulation}

State variable fields of the form $\psi(\mathbf{x})$ are approximated using deep fully-connected neural networks that map the location $\mathbf{x}$ to the corresponding value of $\psi$ at that location. This map can be expressed as $\psi(\mathbf{x}) = \bar{\psi}(\mathbf{x};\boldsymbol{\theta})$, and is defined by the sequence of operations
\begin{subequations}
\begin{align}
\mathbf{z}^0 &= \mathbf{x}, \label{eq:NNInput} \\
\mathbf{z}^k &= \sigma(\mathbf{W}^k \mathbf{z}^{k-1} + \mathbf{b}^k), \quad 1 \le k \le \ell-1, \label{eq:NNMiddleLayers} \\
\psi = \mathbf{z}^\ell &= \mathbf{W}^\ell \mathbf{z}^{\ell-1} + \mathbf{b}^\ell.
\end{align}
\end{subequations}
The input $\mathbf{x}$ is propagated through $\ell$ layers, all of which (except the last) take the form of a linear operation composed with a nonlinear transformation. Each layer outputs a vector $\mathbf{z}^k \in \mathbb{R}^{q_k}$, where $q_k$ is the number of `neurons', and is defined by a weight matrix $\mathbf{W}^k \in \mathbb{R}^{q_k \times q_{k-1}}$, a bias vector $\mathbf{b}^k \in \mathbb{R}^{q_k}$, and a nonlinear activation function $\sigma(\cdot)$. Finally, the output of the last layer is assigned to $\psi$. The weight matrices and bias vectors, which parametrize the map from $\mathbf{x}$ to $\psi$, form a set of trainable parameters $\boldsymbol{\theta} = \{\mathbf{W}^k,\mathbf{b}^k\}_{k=1}^\ell$.

The choice of the nonlinear activation function $\sigma(\cdot)$ and the initialization procedure for the trainable parameters $\boldsymbol{\theta}$ are both important factors in determining the performance of neural networks. While the tanh function has been a popular candidate in the context of PINNs \citep{lu2021a}, recent works have shown that using sinusoidal activation functions can lead to improved training performance by promoting the emergence of small-scale features \cite{sitzmann2020,wong2022}. In this work, we select the sinusoidal representation network (SIREN) architecture from Ref.~\cite{sitzmann2020}, which combines the use of the sine as an activation function with a specific way to initialize the trainable parameters $\boldsymbol{\theta}$ that ensures that the distribution of the input to each sine activation function remains unchanged over successive layers. Specifically, each component of  $\mathbf{W}^k$ is uniformly distributed between $- \sqrt{6/q_k}$ and $\sqrt{6/q_k}$ where $q_k$ is the number of neurons in layer $k$, and $\mathbf{b}^k = \mathbf{0}$, for $k=1, \dots, \ell$. Further, the first layer of the SIREN architecture is $\mathbf{z}^1 = \sigma(\omega_0 \mathbf{W}^1 \mathbf{z}^0 + \mathbf{b}^1)$ instead of \eqref{eq:NNMiddleLayers}, with the extra scalar $\omega_0$ promoting higher-frequency content in the output.

\textcolor{black}{We use the standard logistic function for the sigmoid function representing the material density field; that is, $\rho = \mathrm{sigmoid}(\phi/\delta) = 1/(1+\exp (-\phi/\delta))$, where $\phi = \bar{\phi}(\mathbf{x}; \boldsymbol{\theta}_\phi)$ is the output of the level-set neural network. Note that the presence of the sigmoid function means that $\rho$ can only get asymptotically close to the theoretical solution of the inverse problem, which is not a problem in practice as exemplified by the widespread adoption of sigmoid functions to learn bounded functions in machine learning \cite{goodfellow2016}.}

\subsection{Training procedure}

We calculate the loss and train the neural networks in TensorFlow 2. The training is performed using ADAM, a first-order gradient-descent-based algorithm with adaptive step size \cite{kingma2014}. In each case, we repeat the training over four random initializations of the neural networks parameters and report the best results. Three tricks resulted in noticeably improved training performance and consistency.

First, we found that pretraining the level-set neural network $\phi(\mathbf{x}) = \bar{\phi}(\mathbf{x}; \boldsymbol{\theta}_\phi)$ in a standard supervised setting leads to much more consistent results over different initializations of the neural networks. During this pretraining phase, carried out before the main optimization, we minimize the mean-square error 
\begin{equation}
\mathcal{L}_\mathrm{sup}(\boldsymbol{\theta}_\phi) = \frac{1}{|\Omega^d|} \sum_{\mathbf{x}_i \in \Omega^d} |\bar{\phi}(\mathbf{x}_i; \boldsymbol{\theta}_\phi) - \phi_i |,
\end{equation}
where $\Omega^d$ is the same set of collocation points as in Eq.~(5), the supervised labels $\phi_i = |\mathbf{x}_i| - 0.25$ for the elastic matrix, and $\phi_i = y_i + 0.25$ for the elastic layer. The material density $\bar{\rho}(\mathbf{x};\boldsymbol{\theta}_\phi) = \mathrm{sigmoid}(\bar{\phi}(\mathbf{x};\boldsymbol{\theta}_\phi)/\delta)$ obtained at the end of this pretraining phase is one outside a circle of radius 0.25 centered at the origin for the elastic matrix, and it is one above the horizontal line $y = -0.25$ for the elastic layer. This choice for the supervised labels is justified by the fact that $\rho$ is known to be one along the outer boundary of the domain $\Omega$ for the elastic matrix, and it is known to be one (zero) along the top (bottom) boundary of $\Omega$ for the elastic layer. 

Second, during the main optimization in which all neural networks are trained to minimize the total loss described in Eq.~(3) of the main text, we evaluate the loss component $\mathcal{L}_\mathrm{gov}$ in Eq.~(5) using a different subset, or mini-batch, of residual points from $\Omega^d$ at every iteration. Such a mini-batching approach has been reported to improve the convergence of the PINN training process \citep{wight2021,daw2022}, corroborating our own observations. Thus, each gradient update
\begin{subequations}
\begin{align}
\boldsymbol{\theta}_{\boldsymbol{\psi}}^{k+1} &= \boldsymbol{\theta}_{\boldsymbol{\psi}}^k - \alpha_{\boldsymbol{\psi}}(k) \nabla_{\boldsymbol{\theta}_{\boldsymbol{\psi}}} \mathcal{L}(\boldsymbol{\theta}_{\boldsymbol{\psi}}^k, \boldsymbol{\theta}_\phi^k), \\
\boldsymbol{\theta}_\phi^{k+1} &= \boldsymbol{\theta}_\phi^k - \alpha_\phi(k) \nabla_{\boldsymbol{\theta}_\phi} \mathcal{L}(\boldsymbol{\theta}_{\boldsymbol{\psi}}^k, \boldsymbol{\theta}_\phi^k)
\end{align} \label{eq:GradientUpdate}%
\end{subequations}
is evaluated using a subsequent mini-batch of residual points. The order of the mini-batches is shuffled after every epoch, i.e.~after passing through all the mini-batches constituting $\Omega^d$.  

Third, the initial nominal step size $\alpha_{\boldsymbol{\psi}}$ governing the learning rate of the physical quantities neural networks is set to be 10 times larger than its counterpart $\alpha_\phi$ governing the learning rate of the level-set neural network. This results in a separation of time scales between the rate of change of the physical quantities neural networks and that of the level-set neural network, which is motivated by the idea that physical quantities should be given time to adapt to a given geometry before the geometry itself changes.

\subsection{Parameter values}

In this study, the hyperparameters were manually optimized, starting from a base configuration and changing individual values in order for the training to avoid being stuck in bad local minima. We performed this process once for cases with the same physics and outer matrix geometry, leading to the same hyperparameter values among these cases. Note that there also exist systematic hyperparameter optimization methods, although we have not explored them \citep{akiba2019,yang2020}.

For all cases in Tabs.~I to III except cases 20, 21, 22, we opted for neural networks with 4 hidden layers of 50 neurons each, which we found to be a good compromise between expressivity and training time. For cases 20 and 22, we used 6 hidden layers with 100 neurons each. For case 21, we used 6 hidden layers with 150 neurons each. Further, we choose $\omega_0 = 10$ as the scalar appearing in the first layer of the SIREN architecture.

For all cases in Tabs.~I to III except cases 21 and 22, we considered measurements distributed over 100 equally-spaced locations along each straight external boundary segment (i.e., amounting to $|\partial \Omega^m| = 400$ in cases where measurements were taken along all sides of a square matrix, and $|\partial \Omega^m| = 100$ in the elastic layer cases). For case 21, we considered measurements distributed over 357 locations along the wavy boundary of the half-star. For case 22, we considered measurements distributed over 400 locations in a grid on each face of the 3D cube.

For all cases in Tabs.~I to III except cases 20, 21, 22, the set of collocation points $\Omega^d$ consists of 10000 points. For cases 20 and 22, $\Omega^d$ consists of 50000 points. For case 21, $\Omega^d$ consists of approximately 200000 points. In all cases, the collocation points are sampled from $\Omega$ using a Latin Hypercube Sampling (LHS) strategy.

The pretraining of the level-set neural network is carried out using the Adam optimizer with nominal step size $10^{-3}$ over 800 iterations, employing the whole set $\Omega^d$ to compute the gradient of $\mathcal{L}_\mathrm{sup}$ at each update step. The main optimization, during which all neural networks are trained to minimize the total loss described in Eq.~(3), is also carried out using the Adam optimizer with a mini-batch size of 1000 for all examples. For cases 1 to 19 in Tab.~I and cases 27, 28 in Tab.~III, we use a total of 1500k training iterations starting from a nominal step size of $10^{-3}$, reduced to $10^{-4}$ and $10^{-5}$ at 600k and 1200k iterations, respectively. The schedule is the same for cases 23 to 25 in Tab.~II, with the difference that we use a total of 2000k training iterations. For cases 20 and 22 in Tab.~I, we use a total of 2500k iterations starting from a nominal step size of $10^{-3}$, which is reduced to $10^{-4}$ and $10^{-5}$ at 800k and 2000k iterations, respectively. For case 21 in Tab.~I, we use a total of 2000k iterations starting from a nominal step size of $10^{-3}$, which is reduced to $10^{-4}$ and $10^{-5}$ at 600k and 1200k iterations, respectively. For case 26 in Tab.~III, we use a total of 3000k iterations starting from a nominal step size of $10^{-3}$, which is reduced to $10^{-4}$ and $10^{-5}$ at 600k and 2600k iterations, respectively. As mentioned in the previous section, the initial nominal step size for the level-set neural network is set to $10^{-4}$ in all examples, and decreases to $10^{-5}$ at the same time as the other neural networks. 

The scalar weights in the loss (3) in the main text are assigned the values $\lambda_\mathrm{meas} = 10$, $\lambda_\mathrm{gov} = 1$, and $\lambda_\mathrm{reg} = 1$ for all cases. We also multiply the second term of $\mathcal{L}_\mathrm{gov}$ in \eqref{eq:NonlinearLF} and \eqref{eq:LinearLF} of the main text with a scalar weight $\lambda_\mathrm{cr} = 10$.

Finally, the computations are carried out on a cluster, using 24 threads of an Intel Xeon
Platinum 8260 CPU for each case. For each case, the pretraining phase takes a couple minutes, while the main optimization takes between 2 and 6 hours.

{\color{black}
\section{Effect of measurement noise}
\label{app:EffectMeasurementNoise}

In this appendix, we analyze the robustness of solutions provided by our TO framework in the presence of measurement noise, focusing on mechanical experiments on a linear elastic matrix containing voids (cases 4, 9, 18 in Tab.~I). To this effect, we simulate noisy measurements by adding white Gaussian noise of standard deviation $\sigma_\mathrm{noise}$ to the displacement data obtained from Abaqus simulations. 

Our TO framework identifies the correct structures for noise levels $\sigma_\mathrm{noise}$ up to 10\% of the characteristic displacement magnitude $P_0 L/E$, where $P_0$ is the applied normal traction, $L$ the size of the square matrix, and $E$ the Young's modulus (Supplementary Fig.~\ref{fig:NCS_Noise}a). 

For a more quantitative analysis of the effect of measurement noise on the accuracy of the identified voids, consider an accuracy metric such as the IoU (intersection over union, a non-negative scalar equal to 1 in the perfect case). In general, we can expect the IoU to depend on $\sigma_\mathrm{noise}$, the geometry of the matrix and voids, and the material properties of the matrix. Since the critical quantity affecting the accuracy of the results is the signal-to-noise ratio, we may write the IoU as
\begin{equation}
\mathrm{IoU} = f \left( \frac{\sigma_\mathrm{noise}}{u_\mathrm{meas}^*}, \mathcal{G} \right),
\label{eq:NoiseScaling}
\end{equation}
where $f$ is some unknown function, $u_\mathrm{meas}^*$ is an estimate for the signal amplitude, that is, the average magnitude of boundary displacement perturbation caused by the voids, and $\mathcal{G}$ is a scalar quantity that depends on the normalized geometric shape of the voids, such as their aspect ratio or boundary curvature. The intuition behind including $\mathcal{G}$ is that more complex void shapes are harder to detect accurately than simpler shapes of similar size, even though both induce boundary displacement perturbations of comparable magnitude. To derive an estimate for $u_\mathrm{meas}^*$, consider a circle-shaped (in 2D) or spherical-shaped (in 3D) stress-free void centered inside a disk (in 2D) or a sphere (in 3D) subject to a normal traction $P_0$ on its outer boundary. In this case, it is well-known that the boundary displacement perturbation induced by the presence of the void scales as $(P_0 L/E) (a/L)^d$, where $a$ is the void diameter and $d$ is the spatial dimension (2 in 2D, 3 in 3D) \cite{bower2009}. Thus, we choose $u_\mathrm{meas}^* = (P_0 L/E) (a/L)^d$, with $a$ equal to the average equivalent size of all voids in the matrix, the equivalent size of a given void being defined as the size of a circle-shaped (in 2D) or spherical-shaped (in 3D) void with identical area or volume. 

For all three void configurations tested above, the IoU plotted as a function of $\sigma_\mathrm{noise}/u_\mathrm{meas}^*$ remains close to 1 for low $\sigma_\mathrm{noise}/u_\mathrm{meas}^*$ and starts to decrease noticeably around $\sigma_\mathrm{noise}/u_\mathrm{meas}^* = 1$ (Supplementary Fig.~\ref{fig:NCS_Noise}b). The faster decrease of the IoU for the slit-shaped void may be attributable to its high aspect ratio, which is accounted for by the geometric parameter $\mathcal{G}$ in \eqref{eq:NoiseScaling}. 

In practice, the measurement noise level $\sigma_\mathrm{noise}$ can often be known ahead of time (from the sensor specifications), but $u_\mathrm{meas}^*$ and $\mathcal{G}$ are unknown since the voids are hidden. Nevertheless, we can derive two useful guidelines from the scaling relationship \eqref{eq:NoiseScaling}. First, increasing $P_0$ will always reduce the ratio $\sigma_\mathrm{noise}/u_\mathrm{meas}^*$, leading to better results for any noise level and any void configuration. Second, since the accuracy of the identified voids decreases sharply for $\sigma_\mathrm{noise}/u_\mathrm{meas}^*$ on the order of 1 or greater (Supplementary Fig.~\ref{fig:NCS_Noise}b), we can expect that, for a given $\sigma_\mathrm{noise}$, the minimum detectable void size will be on the order of $(\sigma_\mathrm{noise} E / P_0 L)^{1/d} L$.  This minimum detectable void formula is a worst-case estimate in the limit of otherwise refined numerics; it assumes that voids are situated far from the outer boundaries while numerical parameters such as the interfacial thickness $\delta$ and collocation point separation are sufficiently small compared to the void size, and that the optimizer is perfect.}

\begin{figure*}
\centering
\includegraphics[width=\textwidth]{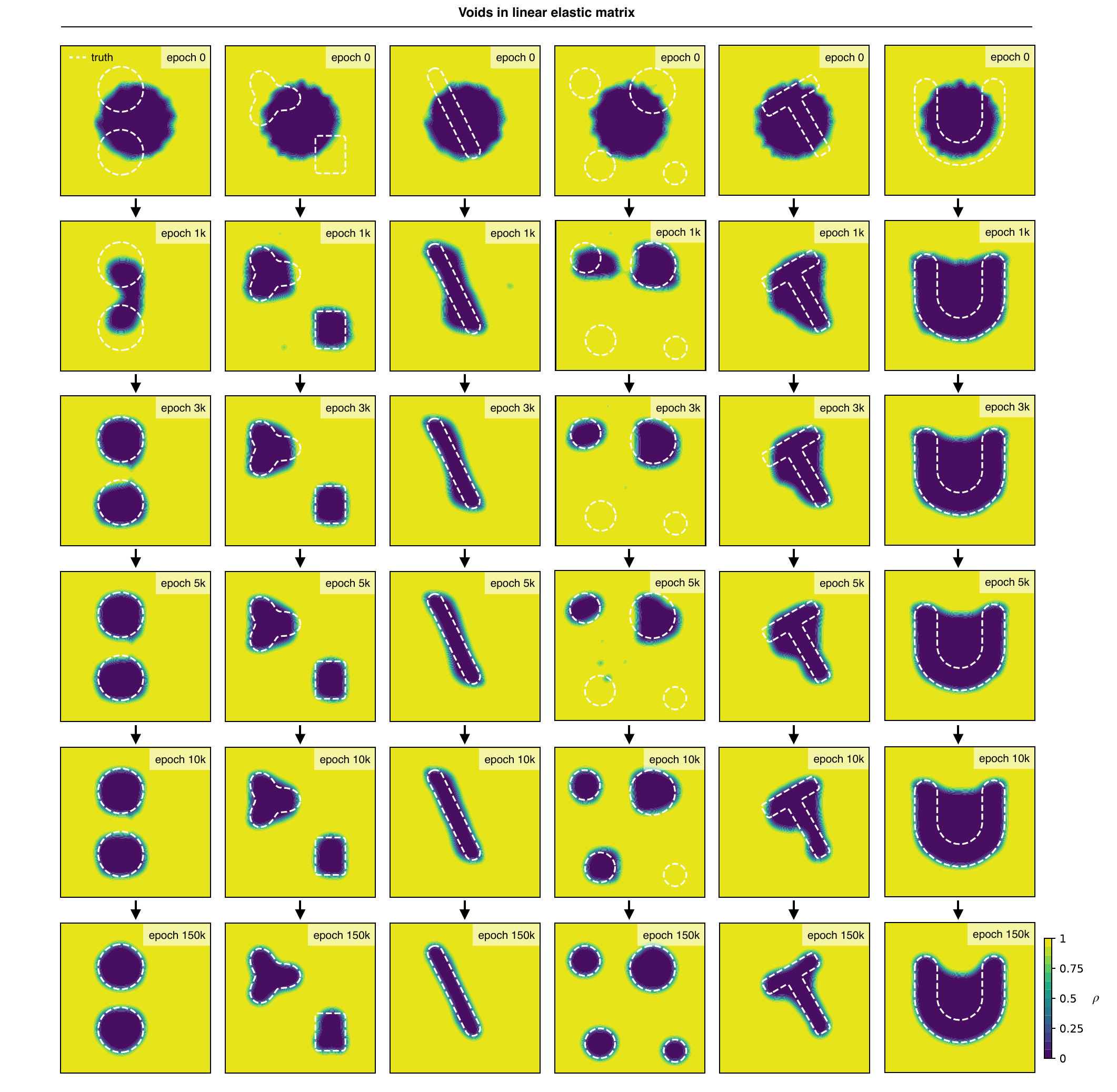}
\caption{\textbf{Identification of voids in a linear elastic matrix.} Evolution of the material density during the training process for the cases reported in Fig.~4a-d.}
\label{fig:NCS_ElaVoid_Iters}
\end{figure*}

\begin{figure*}
\centering
\includegraphics[width=\textwidth]{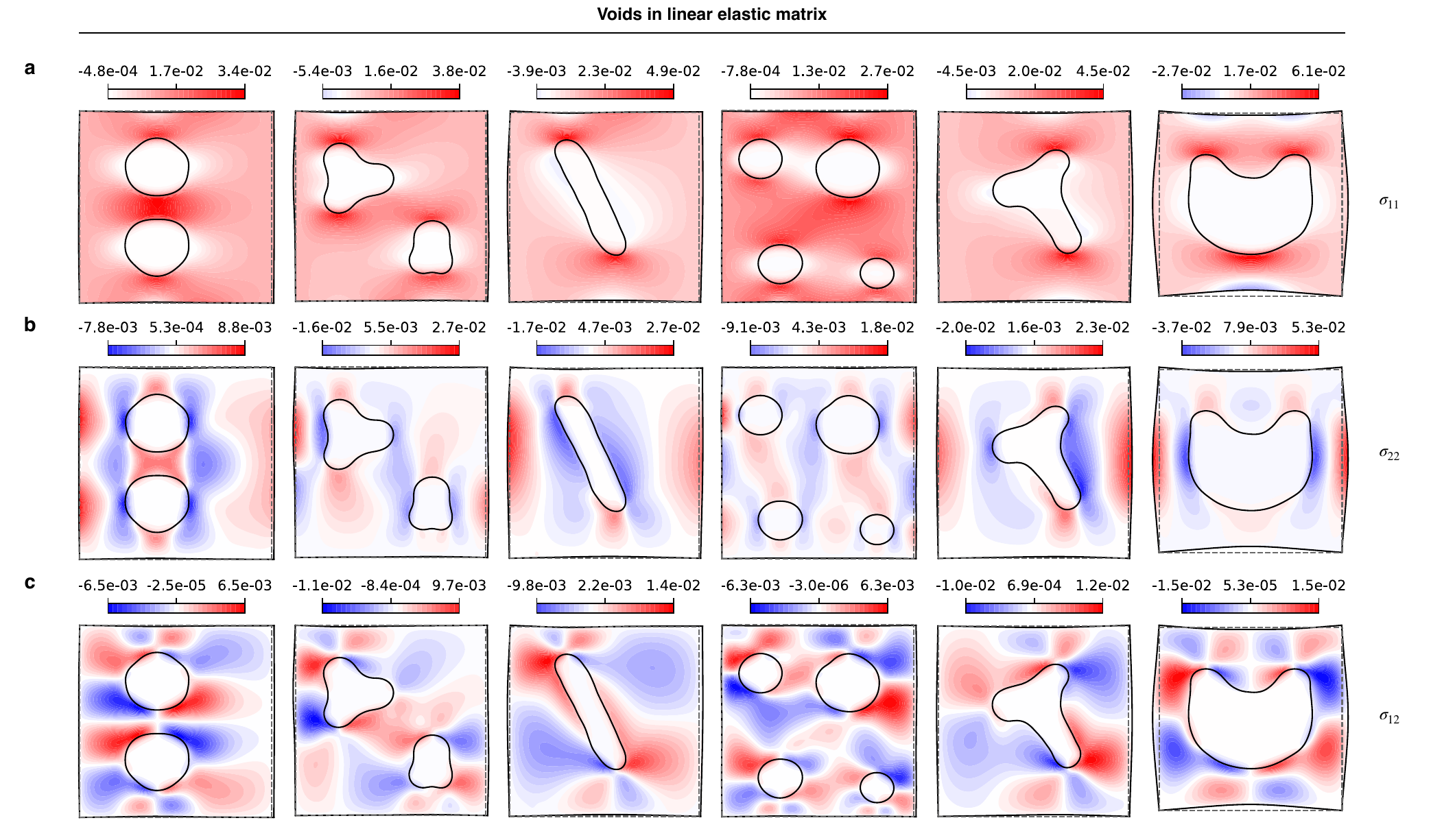}
\caption{\textbf{Identification of voids in a linear elastic matrix.} Final Cauchy stress components $\sigma_{xx}$ (\textbf{a}), $\sigma_{yy}$ (\textbf{b}), and $\sigma_{xy}$ (\textbf{c}), displayed in the deformed configuration obtained from the final displacement components $u_1$ and $u_2$, for the cases reported in Fig.~4a-d. The grey dotted lines show the outline of the matrix surface in the reference configuration.}
\label{fig:NCS_ElaVoid_Stress}
\end{figure*}

\begin{figure*}
\centering
\includegraphics[width=\textwidth]{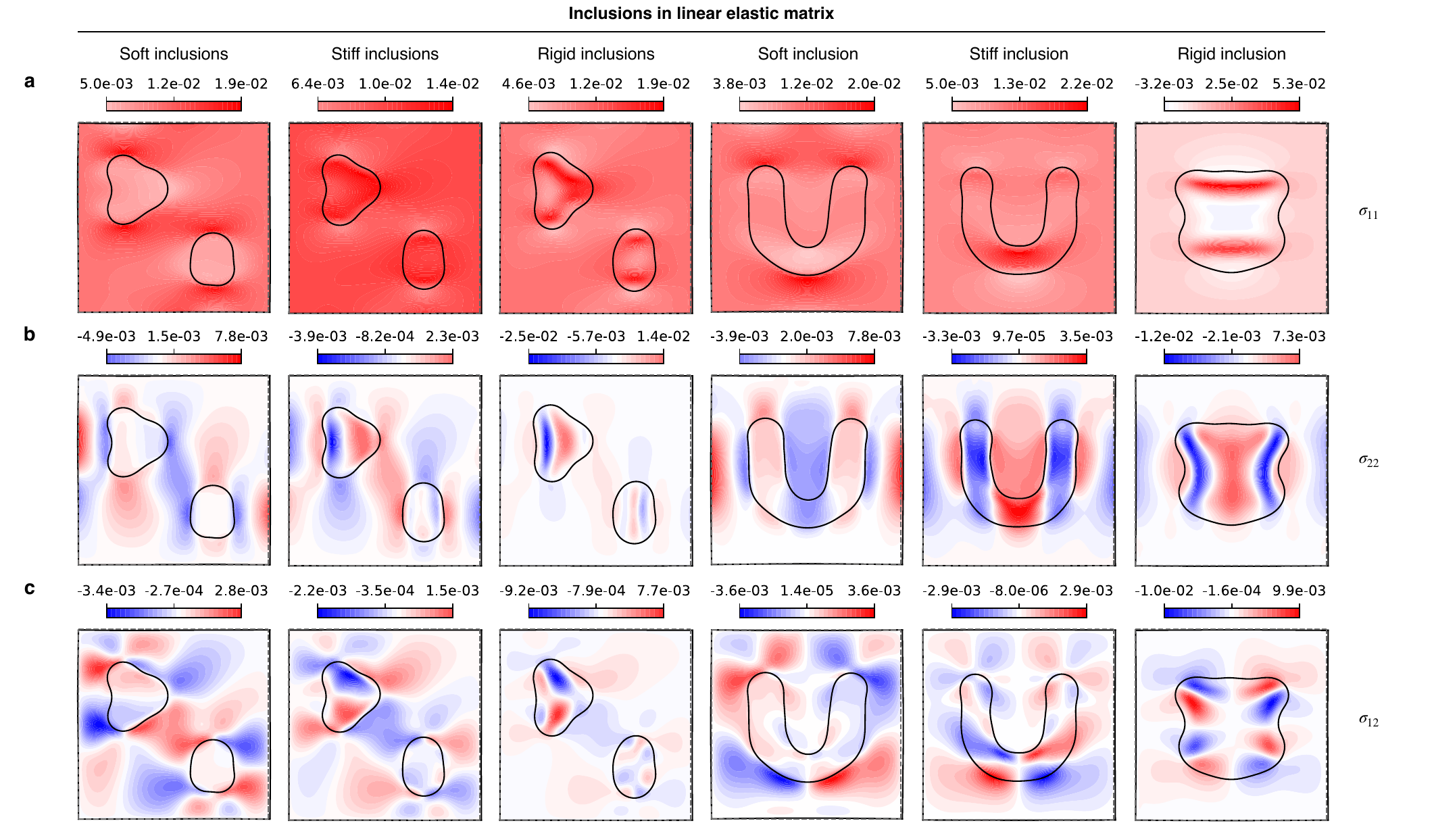}
\caption{\textbf{Identification of inclusions in a linear elastic matrix.} Final Cauchy stress components $\sigma_{xx}$ (\textbf{a}), $\sigma_{yy}$ (\textbf{b}), and $\sigma_{xy}$ (\textbf{c}), displayed in the deformed configuration obtained from the final displacement components $u_1$ and $u_2$, for the cases reported in Fig.~4e. The grey dotted lines show the outline of the matrix surface in the reference configuration.}
\label{fig:NCS_ElaInclusion_Stress}
\end{figure*}

\begin{figure*}
\centering
\includegraphics[width=\textwidth]{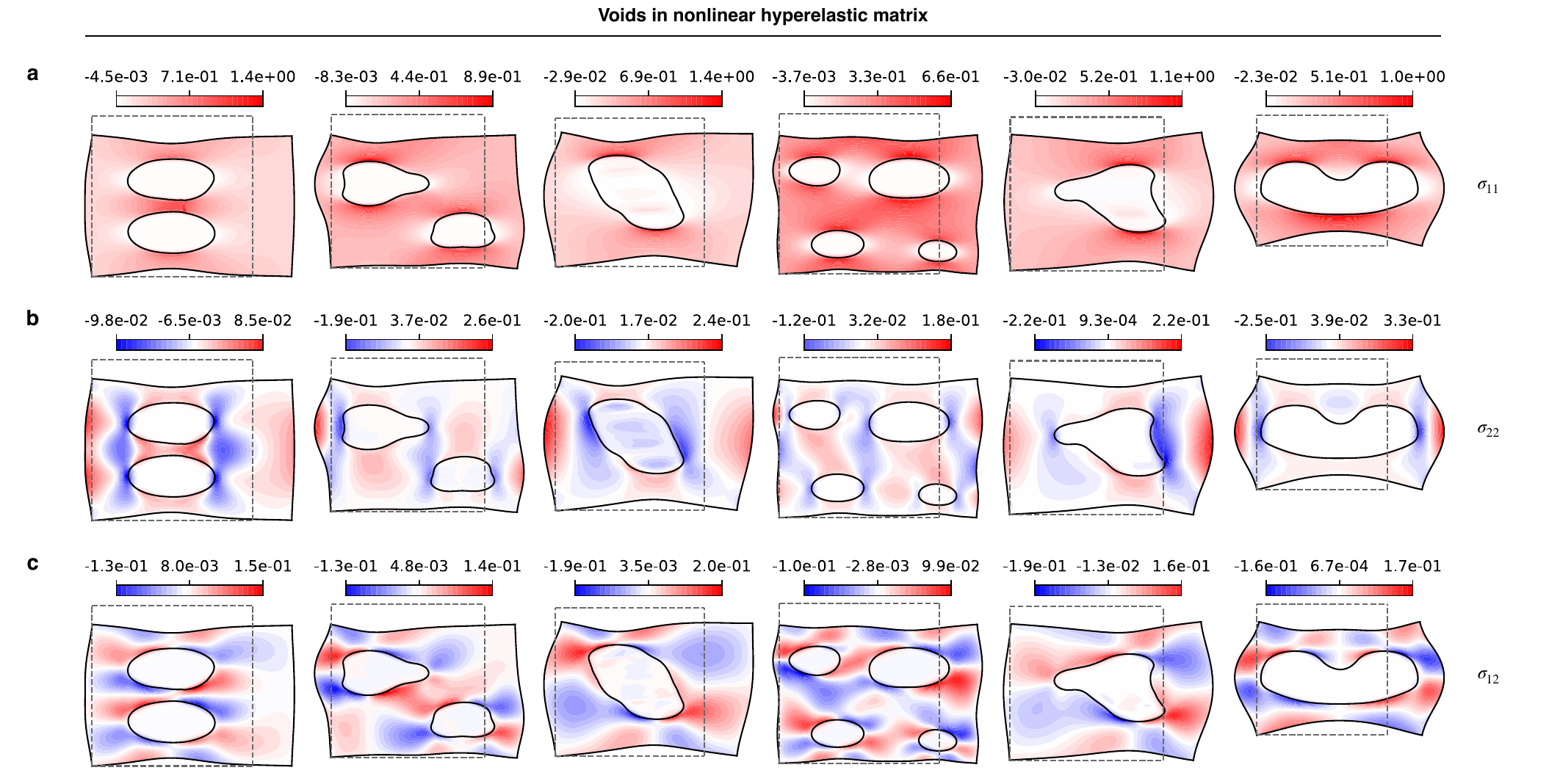}
\caption{\textbf{Identification of voids in a nonlinear hyperelastic matrix.} Final Cauchy stress components $\sigma_{xx}$ (\textbf{a}), $\sigma_{yy}$ (\textbf{b}), and $\sigma_{xy}$ (\textbf{c}), displayed in the deformed configuration obtained from the final displacement components $u_1$ and $u_2$, for the cases reported in Fig.~4f. The grey dotted lines show the outline of the matrix surface in the reference configuration.}
\label{fig:NCS_HyperElaVoid_Stress}
\end{figure*}

\begin{figure*}
\centering
\includegraphics[width=\textwidth]{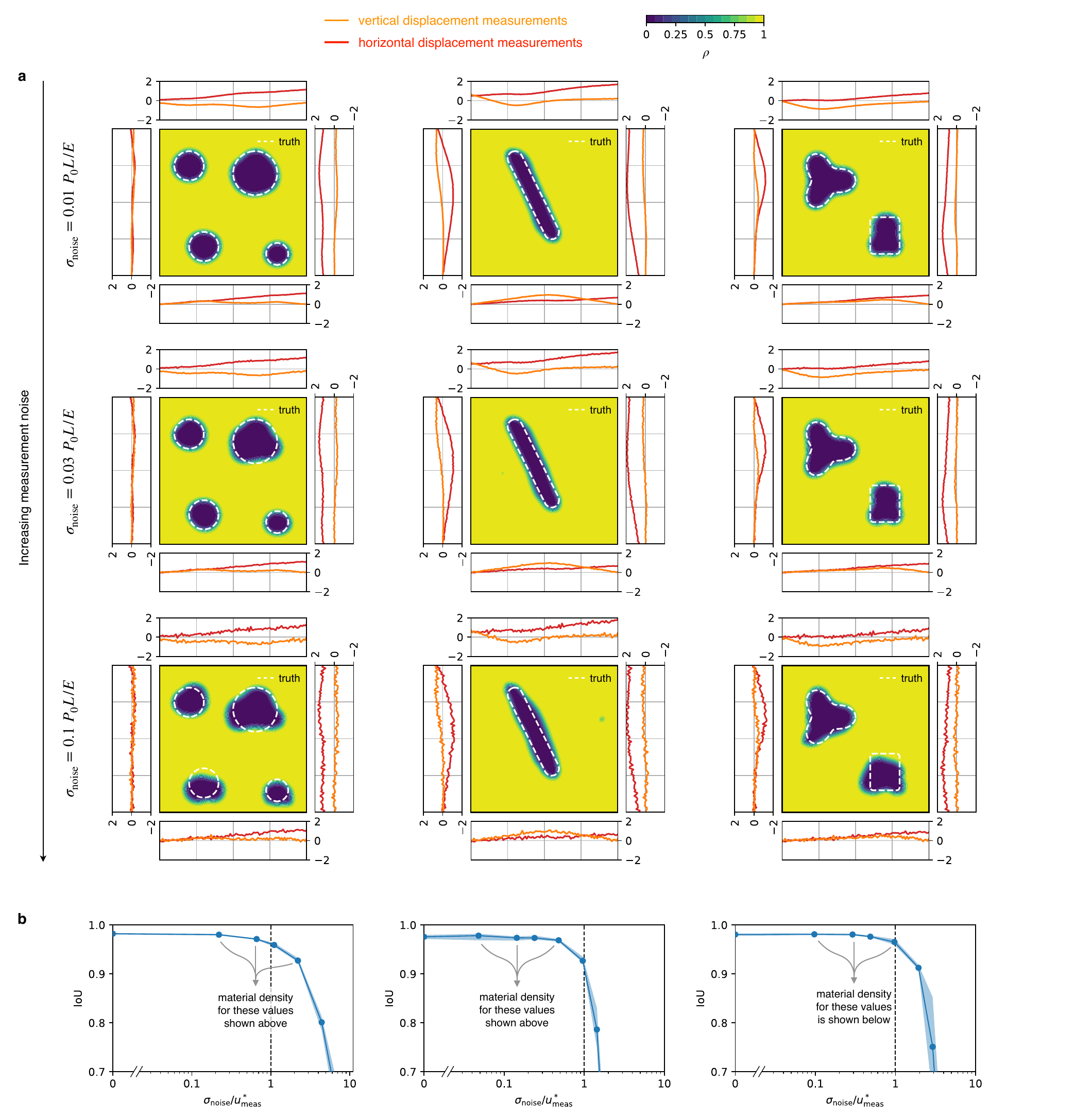}
\caption{\textbf{Effect of noise on the identification of voids in a linear elastic matrix.} \textbf{a}, Final material density $\rho$ obtained for various values of measurement noise standard deviation $\sigma_\mathrm{noise}$. \textbf{b}, IoU (intersection over union, a geometry detection accuracy metric equal to 1 in the perfect case) versus scaled standard deviation $\sigma_\mathrm{noise}/u_\mathrm{meas}^*$ of the measurement noise, where $u_\mathrm{meas}^*$ is an analytical estimate for the magnitude of boundary displacement perturbation caused by the voids based on their average size as well as the size and material properties of the elastic matrix. Each plot corresponds to the voids topology shown in \textbf{a} in the same column. The circles report the average value and the shade report the highest and lowest values obtained for 4 random initializations of the neural networks parameters.}
\label{fig:NCS_Noise}
\end{figure*}

\begin{figure*}
\centering
\includegraphics[width=\textwidth]{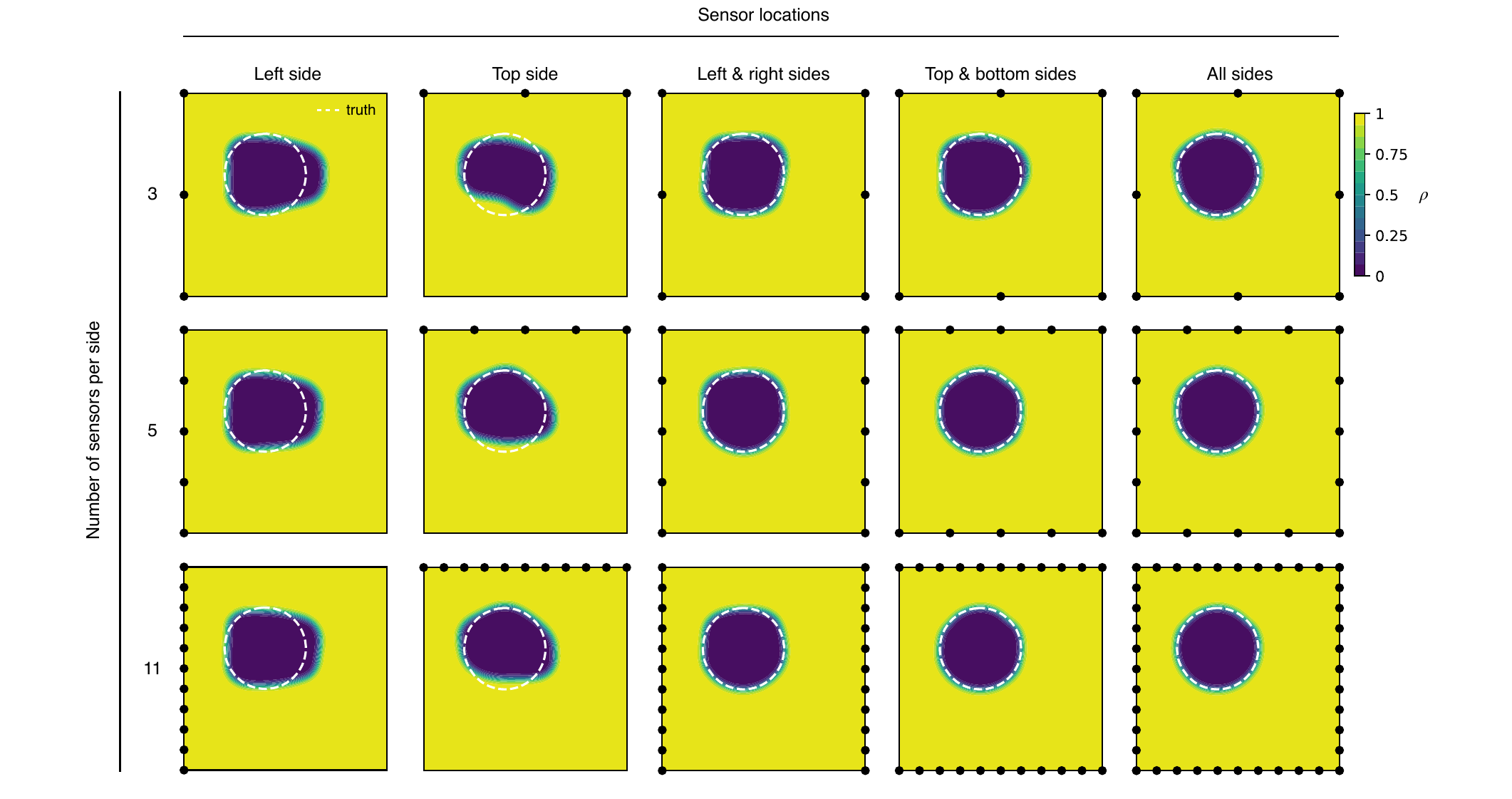}
\caption{\textbf{Effect of sparse measurements on the identification of voids in a linear elastic matrix with one circle-shaped void.} Final material density $\rho$ obtained in our framework when using fewer measurement locations and restricting the number of measurements to a subset of the outer surfaces.}
\label{fig:NCS_SparseMeasCircle}
\end{figure*}

\begin{figure*}
\centering
\includegraphics[width=\textwidth]{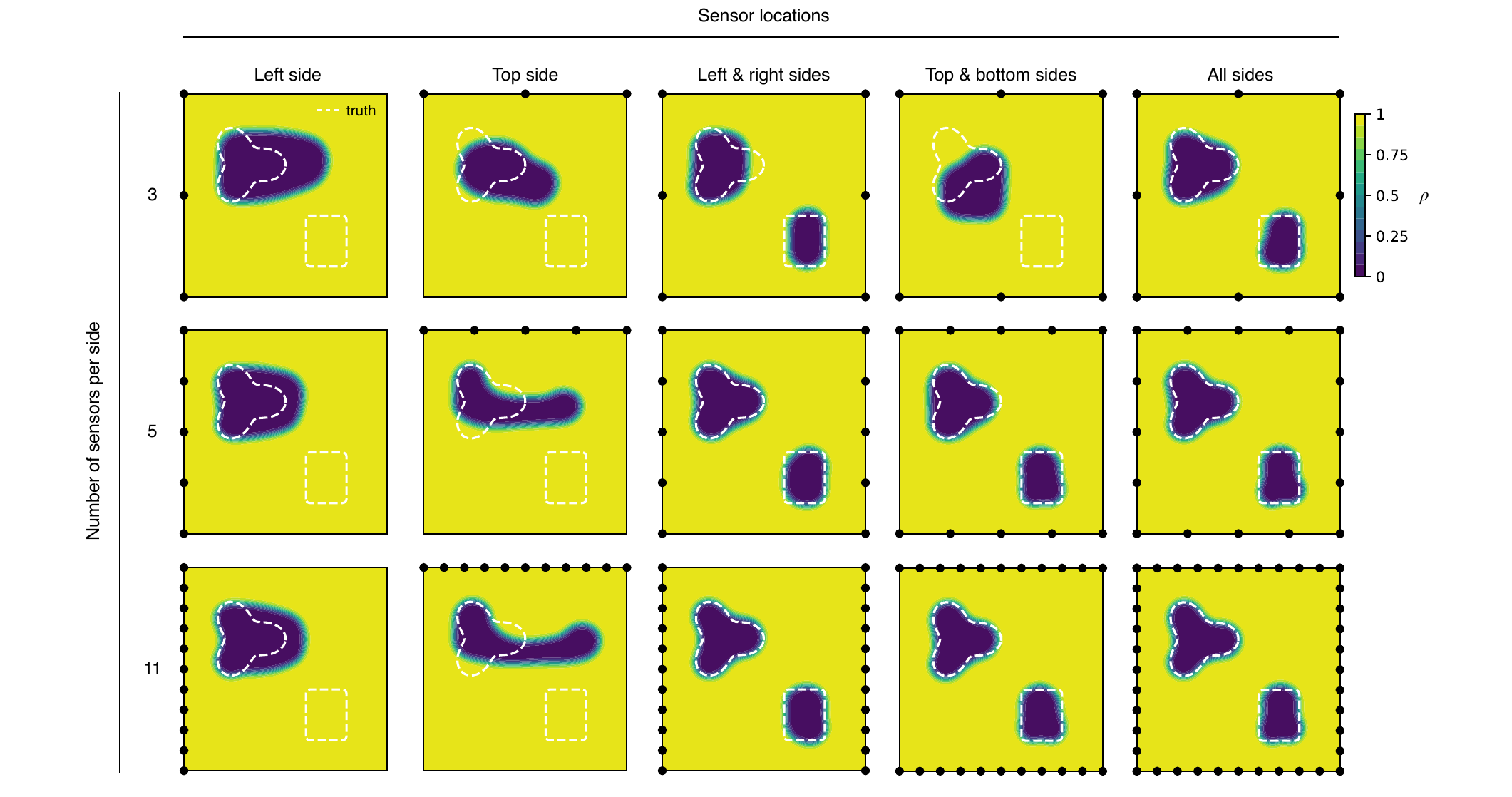}
\caption{\textbf{Effect of sparse measurements on the identification of voids in a linear elastic matrix with one star-shaped and one rectangle-shaped void.} Final material density $\rho$ obtained in our framework when using fewer measurement locations and restricting the number of measurements to a subset of the outer surfaces.}
\label{fig:NCS_SparseMeasTwo}
\end{figure*}

\begin{figure*}
\centering
\includegraphics[width=\textwidth]{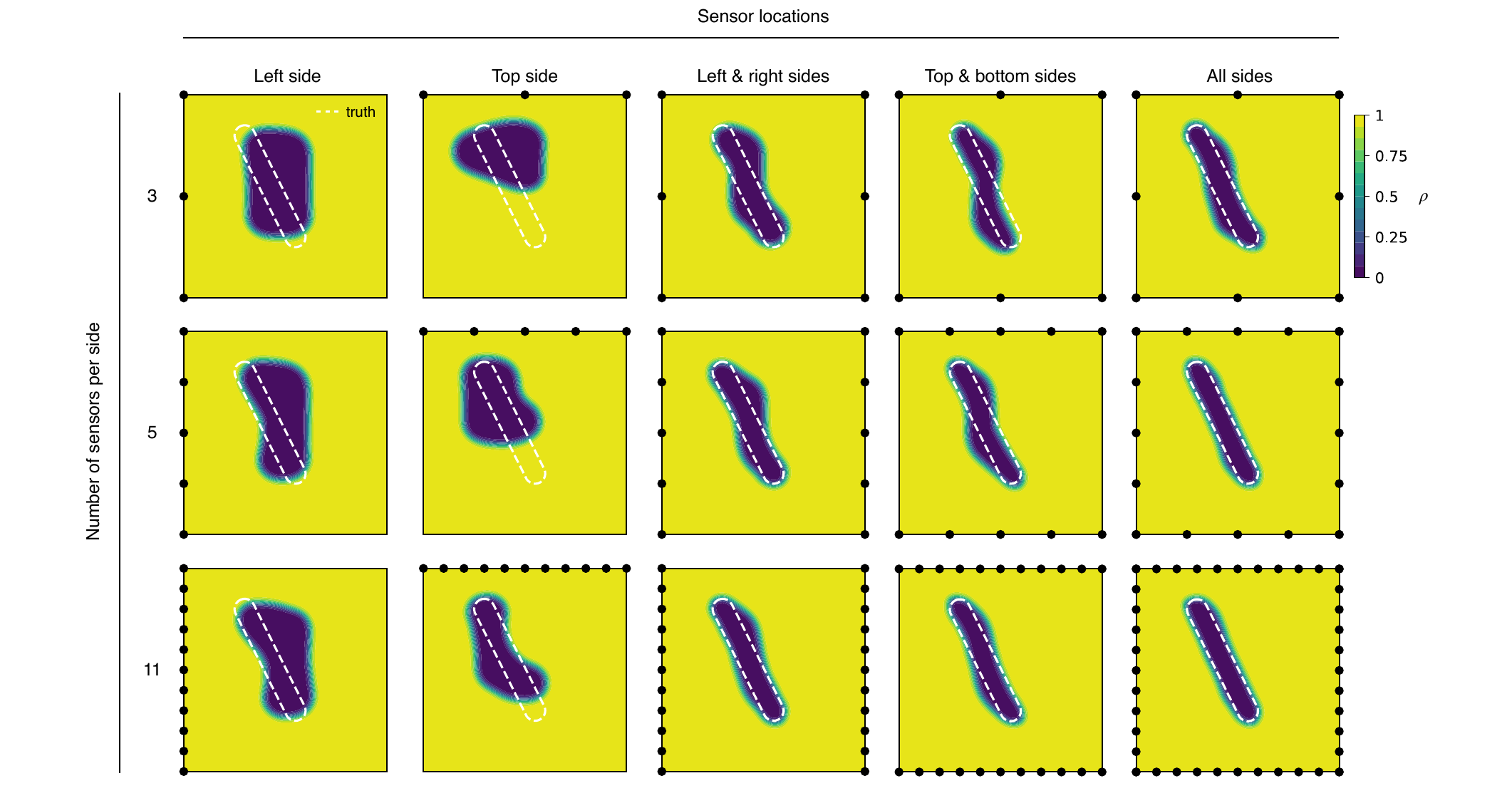}
\caption{\textbf{Effect of sparse measurements on the identification of voids in a linear elastic matrix with one slit-shaped void.} Final material density $\rho$ obtained in our framework when using fewer measurement locations and restricting the number of measurements to a subset of the outer surfaces.}
\label{fig:NCS_SparseMeasSlit}
\end{figure*}

\begin{figure*}
\centering
\includegraphics[width=\textwidth]{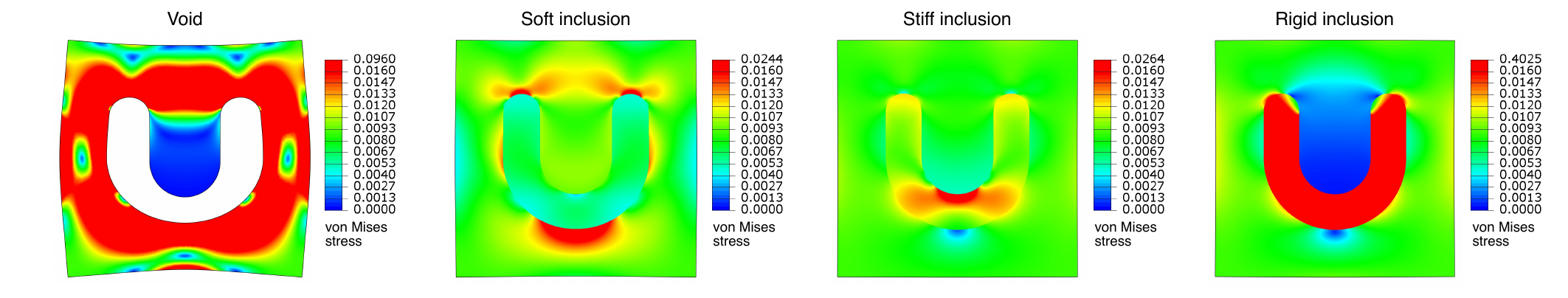}
\caption{\textbf{Stress distribution in a linear elastic matrix for different types of U-shaped inclusions.} The von Mises stress obtained in Abaqus for U-shaped inclusions with different constitutive properties reveals that the concave part of the matrix is subject to very little stress in the case of a void or rigid inclusion. Note also that stiff and rigid inclusions `strengthen' the matrix, as opposed to the void and soft inclusion that `soften' the matrix.}
\label{fig:NCS_AbaqusStress}
\end{figure*}

\begin{figure}
\centering
\includegraphics[width=\textwidth]{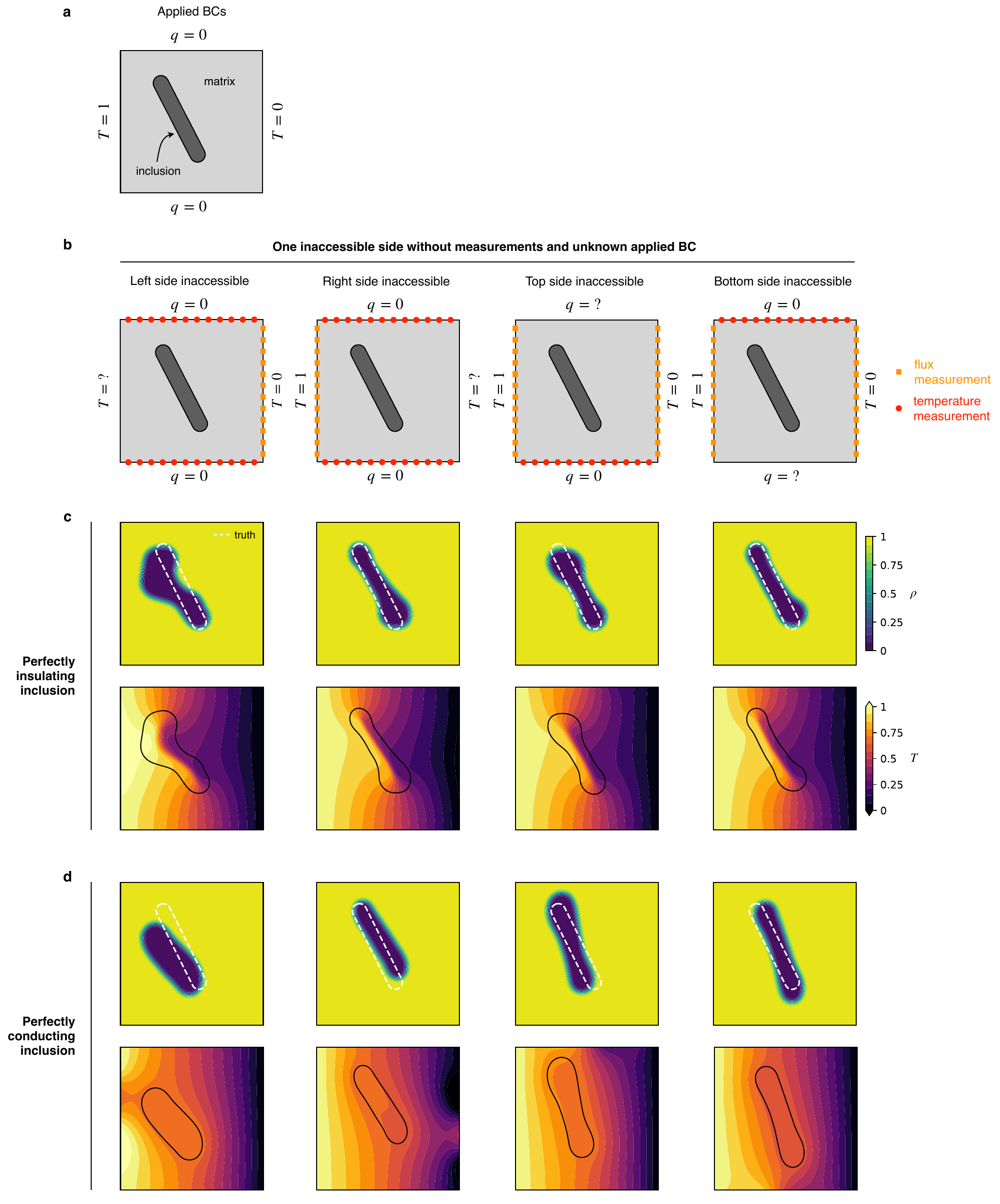}
\caption{\textbf{Identification of inclusions in a nonlinearly conducting matrix with incomplete information.} \textbf{a}, Setup of the geometry and applied boundary conditions during the loading. \textbf{b}, The geometry identification inverse problem is solved assuming that one of the four sides of the matrix is inaccessible to the user, meaning that both the applied boundary condition and the measurements are unavailable on that side. \textbf{c}, The final inferred material density $\rho$ and temperature $T$ in the case of a perfectly insulating inclusion. \textbf{d}, The final inferred material density $\rho$ and temperature $T$ in the case of a perfectly conducting inclusion.}
\label{fig:HeatTransfer}
\end{figure}

\bibliography{bibliography}